\documentclass[times,preprint,3p]{elsarticle}

\journal{Applied Numerical Mathematics}
\usepackage{framed,multirow}
\usepackage{float}

\usepackage{bm}
\usepackage{multirow}
\usepackage{hyperref}
\usepackage{color}
\usepackage{amsfonts}
\usepackage{amsmath}
\usepackage{float}
\usepackage{graphicx}
\usepackage{rotating}
\usepackage{mathtools}
\usepackage[shortlabels]{enumitem}
\usepackage{siunitx} % needed for printing a scientific number \num{e-10}
\usepackage{graphicx}
\usepackage{subcaption}
\usepackage{hhline}
\usepackage{hyperref}
\usepackage[capitalise]{cleveref}

\hypersetup{
    colorlinks=true,
    linkcolor=blue,
    filecolor=magenta,      
    urlcolor=cyan,
}

\usepackage{booktabs}

\newcommand{\bff}{\mathbf{f}}
\newcommand{\bfg}{\mathbf{g}}
\newcommand{\bfk}{\mathbf{k}}

\newcommand{\bfx}{\mathbf{x}}

\newcommand{\bfC}{\mathbf{C}}

\newcommand{\bfK}{\mathbf{K}}

\usepackage{comment}

\renewcommand{\vec}[1]{\mathbf{#1}}

\newcommand{\mm}[1]{\rm mm}

\newcommand{\beq}{\begin{equation}}
\newcommand{\eeq}{\end{equation}}
\newcommand{\bea}{\begin{eqnarray}}
\newcommand{\eea}{\end{eqnarray}}

\newcommand{\bit}{\begin{itemize}}
\newcommand{\eit}{\end{itemize}}
\newcommand{\ben}{\begin{enumerate}}
\newcommand{\een}{\end{enumerate}}

\newcommand{\bK}{\mathbf{K}}
\newcommand{\bk}{\mathbf{k}}

\newcommand{\bX}{\mathbf{X}}

\newcommand{\be}{\mathbf{e}}
\newcommand{\bg}{\mathbf{g}}
\newcommand{\bh}{\mathbf{h}}

\newcommand{\btt}{\mathbf{t}}

\newcommand{\bx}{\mathbf{x}}
\newcommand{\by}{\mathbf{y}}
\newcommand{\bz}{\mathbf{z}}

\newcommand{\erf}[1]{\operatorname{erf} \left [ #1 \right ]}
\newcommand{\expo}[1]{\exp \left [ #1 \right ]}

\newcommand{\norm}[1]{\left\lVert#1\right\rVert}

\makeatletter
\def\@xfootnote[#1]{%
  \protected@xdef\@thefnmark{#1}%
  \@footnotemark\@footnotetext}
\makeatother

\usepackage{url}
\usepackage{xcolor}
\definecolor{newcolor}{rgb}{.8,.349,.1}

\begin{document}
\begin{frontmatter}

%%%%%%%%%%%%%%%%%%%%%%%%%%%%%%%%%%%%%%%%%%%%%%%%%%
%% Title
%%%%%%%%%%%%%%%%%%%%%%%%%%%%%%%%%%%%%%%%%%%%%%%%%%
\title{GP-Recipe: Gaussian Process approximation to linear operations in numerical methods}
%using Gaussian Process modeling and a multidimensional optimal order detection algorithm}

%%%%%%%%%%%%%%%%%%%%%%%%%%%%%%%%%%%%%%%%%%%%%%%%%%
%% section : Authors Information
%%%%%%%%%%%%%%%%%%%%%%%%%%%%%%%%%%%%%%%%%%%%%%%%%%
%Chris DeGrendele
\author[1]{Chris DeGrendele}
\ead{cdegrend@ucsc.edu}
%% Dongwook Lee
\author[1]{Dongwook {Lee}\corref{cor1}}
\cortext[cor1]{Corresponding author: Tel.: +1-831-502-7708}
\ead{dlee79@ucsc.edu}

\address[1]{Department of Applied Mathematics, 
The University of California, Santa Cruz, CA, United States}

%%%%%%%%%%%%%%%%%%%%%%%%%%%%%%%%%%%%%%%%%%%%%%%%%%
%% Abstract
%%%%%%%%%%%%%%%%%%%%%%%%%%%%%%%%%%%%%%%%%%%%%%%%%%
\begin{abstract}
We introduce new Gaussian Process (GP) high-order approximations 
to linear operations that are frequently used
in various numerical methods. 
Our method employs the kernel-based GP regression modeling, a non-parametric
Bayesian approach to regression that operates on the probability distribution over all admissible
functions that fit observed data.
We begin in the first part with 
discrete data approximations to various linear operators applied to smooth data
using the most popular squared exponential kernel function.
In the second part, we discuss data interpolation across discontinuities with sharp gradients,
for which we introduce a new GP kernel that fits discontinuous data without oscillations.
The current study extends our previous GP work on polynomial-free 
shock-capturing methods in finite difference and finite volume methods
to a suite of linear operator approximations on smooth data.
The formulations introduced in this paper can be readily adopted in 
daily practices in numerical methods, including numerical approximations of
finite differences, quadrature rules, interpolations, and reconstructions, 
which are most frequently used in numerical modeling in 
modern science and engineering applications.
In the test problems, we demonstrate that the GP approximated solutions
feature improved solution accuracy compared to the conventional
finite-difference counterparts.
\end{abstract}

%%%%%%%%%%%%%%%%%%%%%%%%%%%%%%%%%%%%%%%%%%%%%%%%%%
%% Keywords up to six entries
%%%%%%%%%%%%%%%%%%%%%%%%%%%%%%%%%%%%%%%%%%%%%%%%%%
\end{frontmatter}

%\linenumbers

%%%%%%%%%%%%%%%%%%%%%%%%%%%%%%%
%       Introduction
%%%%%%%%%%%%%%%%%%%%%%%%%%%%%%%
\section{Introduction}
\noindent
Many physical phenomena in science and engineering applications pose numerical representations to function approximations that involve interpolations, derivatives, and integrals. Popular numerical methods for interpolation include polynomial interpolations, spline and Hermite interpolations, and rational function interpolations. For numerical derivatives, finite differences are perhaps the most popular method of all, while other alternatives are spectral methods, spline differentiation, and least square differentiation, to name a few. For integrals, there exist many quadrature rules to numerically evaluate the values of integrated functions. All of the methods mentioned can be broadly classified as \textit{parametric} as they formulate function approximations by assuming a fixed functional form (e.g., polynomial vs. trigonometric functions) and configure relevant function parameters (e.g., polynomial coefficients, polynomial degrees).
The primary aim of this paper is to present various types of function approximations based on the \textit{nonparametric} Gaussian process (GP) modeling approach, where no assumption is made for a specific functional form but rather the function approximation is entirely determined by observed data.
%\cite{lele1992compact}
%

Since its first introduction in \cite{williams1995gaussian}, GP has been employed as a data-driven tool to study data-dominant problems in various applications. The scientific interests in GP regression brought broader attention in the subsequent work \cite{rasmussen2005} by the same authors, where supervised machine learning using GP regression was presented. Prior supervised machine learning models were mostly parametric in nature, which lacked the ability to generalize to complex data sets and often faced data overfitting issues \cite{rasmussen2003gaussian}. Neural networks are a subset of this class of approaches and thus inherited many of these problems. 
By contrast, GP is a nonparametric, kernel-based Bayesian approach to data regression making data-driven predictions. By being nonparametric, GP regression operates directly on defining a probability distribution over \textit{all} admissible functions that fit the data instead of building around the probability distribution of parameters of any \textit{specific} functional form. 

The GP data regression and predictions are formulated by specifying a prior (on a target function space, e.g., smooth vs. discontinuous, periodic, etc.) and calculating the associated updated posterior distribution inferred by the conditional distribution of an unobserved new data point given the observed data points. This gives the ability to compute the predictive posterior distribution on a set of new data points. The underlying principle in GP regression is to establish the relationship between input and output data points by correlating them and letting them determine the inferred data complexity in the sense of Bayesian inference \cite{rasmussen2005}.

Practitioners usually adopt one of two viewpoints to understand GPs: the Bayesian and function-space. Under the Bayesian viewpoint, estimations of function values are linear \cite{hofmann2008kernel} in the constructed function space, referred to as the weight-space view that centers around probability distributions over linear weights of the sampled input data. The function-space view directly defines a distribution over functions, which ultimately arrives at the same results as with the weight-space approach \cite{rasmussen2005,schulz2018tutorial}. The equivalence of the function-space and weight-space approaches is justified by realizing that defining a distribution of functions is identical to defining each set of distributions of weights, where a collection of sets of weights comprises a collection of functions. The function-space view is what we adopt by default in this paper.

The interest in GP has grown over decades rapidly as a powerful tool for modeling correlated datasets in a wide range of application areas, including machine learning \cite{rasmussen2005}, data science \cite{liu2020gaussian,das2018fast}, deep neural networks \cite{matthews2018gaussian,garriga2018deep}, time-series modeling \cite{roberts2013gaussian,foreman2017fast,corani2021time,brahim2004gaussian}, robotics and control systems \cite{deisenroth2013gaussian,kuss2003gaussian}, computer vision \cite{reeves2020gaussian}, medical applications \cite{ziegler2014individualized}, financial modeling and price forecasting \cite{han2016gaussian,xu2023price}, to name a few. Tutorial articles \cite{schulz2018tutorial,williams1998prediction} have also been reported to provide the community with practical introductions of GP regression as a powerful tool for understanding agnostic function approximations.

%%%%%%%%%%%%%%%%%%%%%%%%%%%%%%%
%       Intro Section 1.1
%%%%%%%%%%%%%%%%%%%%%%%%%%%%%%%
\subsection{Previous Studies of GP Regression in Shock-capturing Schemes}
Apart from the above applications, GP regression has been recently extended to a class of high-order methods in finite difference (FD) and finite volume (FV) formalisms. In \cite{lee2017new}, GP's updated posterior mean function was first introduced as a high-order method for a one-dimensional system of Euler's equations. It demonstrated, as a proof-of-concept, that the new GP method can deliver a variable odd-order convergence rate from the 3rd to the 11th accuracy on a smooth Gaussian profile advection. This preliminary study also considered the GP method on the 1D Shu-Osher shock-tube problem \cite{shu1989} where the shock strength is mild. The GP shock-capturing strategy made a leap forward in the subsequent study reported in \cite{reyes2018new}. This work established a foundational ground where GP's updated posterior mean function as a high-order reconstruction method was integrated into WENO's shock-capturing framework \cite{jiang1996efficient}. The GP solution was tested against a suite of 1D shock-tube tests in hydrodynamics and magnetohydrodynamics. An extension to multiple spatial dimensions, dubbed GP-WENO, was first studied in the finite difference method (FDM) framework in \cite{reyes2019variable}, featuring the variable odd-order accuracy on smooth flows and stable shock-capturing capability on discontinuous flows.

The mathematical ideas of the GP-WENO method have been formulated into a more general kernel-based FV reconstruction method in the finite volume method (FVM) framework in \cite{may2024high,may2024kfvm}. Unlike the use of symmetric positive definite kernels in the previous GP studies \cite{lee2017new,reyes2018new,reyes2019variable}  that explored the GP covariance kernel functions, the work in \cite{may2024high,may2024kfvm} utilized non-symmetric kernel functions (that are no longer GP covariance kernels) and achieved the numerical stability near shocks using the adaptive-order WENO approach \cite{balsara_efficient_2016,reyes2018new}. This kernel-based FVM with WENO (or KFVM-WENO) is shown to be at least sixth-order accurate on the smooth nonlinear isentropic vortex problem \cite{shu1998essentially,spiegel2015survey,reyes2019variable} and robust on a selectively chosen set of benchmark problems, including a three-dimensional compressible isothermal turbulence problem and the Taylor-Green vortex problem. Interested readers are encouraged to consult more details on the setups and references in \cite{may2024kfvm}.

A new effort to restructure the shock-capturing paradigm has also been considered in GP. The majority of the modern shock-capturing algorithms take the so-called \textit{a priori} approach, implying that the shock-detection mechanism is to be always engaged for non-oscillatory fluid data approximations. All of the aforementioned schemes in the previous paragraph belong to this class. On the other hand, an alternative method called \textit{a posteriori} approach maintains non-oscillatory solutions near shocks by post-processing its solutions, where the solutions are cascaded to non-oscillatory, positivity-preserving lower-order solutions if unphysical oscillations are detected in shock regions. This order-cascading strategy is referred to as the MOOD (multidimensional optimal order detection) method and was originally proposed and studied in \cite{clain2011high,diot2012improved,diot2013multidimensional,diot2012methode,bourriaud2020priori} for polynomial-based methods. In \cite{lee2020gp,bourgeois2022gp}, the high-order GP method was successfully integrated into the MOOD architecture (replacing the polynomial counterparts). It showed that the new GP-MOOD scheme gains significant computational efficiency by removing the need for expensive nonlinear shock-detecting switches (e.g., slope limiters, smoothness indicators) while delivering high-order solutions ranging the third, fifth, and seventh-order accuracy in two-dimensional finite volume simulations. 
Moreover, the GP scheme has been extended to a third-order accurate prolongation algorithm in the finite volume adaptive mesh refinement simulations \cite{reeves2020application}.

%%%%%%%%%%%%%%%%%%%%%%%%%%%%%%%
%       Intro Section 1.2
%%%%%%%%%%%%%%%%%%%%%%%%%%%%%%%
\subsection{Extending GP's Data Conversion Property to Interpolations and Linear Operators}
One aspect that allows GP regression great flexibility, particularly in high-order FV reconstruction in multiple spatial dimensions, is its ability to convert data to any specified type. As already investigated (at least partly) in the GP shock-capturing methods mentioned above, GP can convert different data types from one grid location to another in a simpler manner than in the existing approaches \cite{mccorquodale2011high,buchmuller2014improved}. 
For example, a standard FV polynomial reconstruction scheme would require taking \textit{volume-averaged} values at cell centers as input to reconstruct \textit{pointwise} Riemann states as output at cell interface locations in 1D. In multiple dimensions, a quadrature-based integration scheme would need to be used to convert these pointwise estimates to a face integral flux value. A popular way to obtain high-order accurate Riemann states in conventional polynomial-based multidimensional reconstruction schemes is to add transverse corrections for each spatial direction dimension-by-dimension. An example in 3D would include three separate reconstructions, cascading from cell averages in 3D to face averages, face averages to line averages, and finally, line averages to pointwise values and vice versa \cite{mccorquodale2011high,buchmuller2014improved}. Although this process is not particularly difficult, there are many steps that can be computationally compounded, and one may need to apply boundary conditions on intermediate quantities \cite{may2024kfvm}, which results in added data communication in parallel computing. 

In GP, however, such data-type conversions can be achieved by applying the needed integral operations directly to the underlying pointwise GP covariance kernels, thanks to the fact that GP is invariant under linear operations \cite{rasmussen2005,lee2017new,reyes2018new,reyes2019variable,reeves2020application,bourgeois2022gp}. This aspect is the centerpiece of the current study, where we apply GP approximation to a collection of linear operations, such as derivatives and integrations, which are frequently used in modern science and engineering applications.

%%%%%%%%%%%%%%%%%%%%%%%%%%%%%%%
%       Section 2
%%%%%%%%%%%%%%%%%%%%%%%%%%%%%%%
\section{Gaussian Process for Modeling Functions}
\label{sec:GP_modeling}
In this section, we give a concise outline of the GP regression theory from the function-space perspective to prepare basic building blocks that are useful for the rest of the paper. Interested readers are referred to our previous publications \cite{reyes2018new,reyes2019variable,bourgeois2022gp} for similar introductory discussions or to \cite{rasmussen2005,schulz2018tutorial} for a broader theoretical background of GP.

%%%%%%%%%%%%%%%%%%%%%%%%%%%%%%%
%       Section 2.1
%%%%%%%%%%%%%%%%%%%%%%%%%%%%%%%
\subsection{GP Prior and Posterior Predictions}
\label{sec:GP_prior_posterior}
A GP makes predictions of agnostic function value estimations by defining a probabilistic distribution over all admissible functions that can fit the given data. Constructing a collection of random variables is construed as defining a GP, where a finite number of which follow a joint multivariate Gaussian distribution. We say the function $f$ is distributed as a GP and write:

\beq
\label{eq:GP_def}
f(\bx) \sim \mathcal{GP}\left(m(\bx), K(\bx, \bx')\right),
\eeq
where $\bx$ and $\bx'$ are two points (not necessarily distinct) in $\mathbb{R}^D$.
That any agnostic function $f(\bx)$ is in GP according to \cref{eq:GP_def} implies that the observation of the function evaluation of $f$ at a new point $\bx_* \in \mathbb{R}^D$ (where the true value $f(\bx_*)$ has not been observed) follows a joint multivariate Gaussian distribution, allowing us to make a probabilistically meaningful estimation on $f(\bx_*)$ with a predicted uncertainty.

In a more specific term, assume that we have observed the function values of $f$ at a finite set of points $\{\bx: \bx \in \mathbb{R}^D\}$ probabilistically (but no information is available for the function $f$ itself) in terms of the \textit{prior} GP distribution, where the GP \textit{prior} is determined by the following two functions:
\begin{itemize}
\item a mean function $m(\bx)=\bar{f}(\mathbf{x}) = \mathbb{E}[f(\bfx)]$, which is the expected function value of $f$ at the input $\bx \in \mathbb{R}^D$, and 
\item a symmetric positive definite covariance GP kernel function $K(\mathbf{x},\mathbf{x}') =
%\left\langle 
\mathbb{E}\left[
\left(f(\mathbf{x})-\bar{f}(\mathbf{x})\right)
\left(f(\mathbf{x}')-\bar{f}(\mathbf{x}')\right)
%\right\rangle 
\right]$,
which correlates the dependence between the function values at two input points (not necessarily distinct) $\bx$ and $\bx'$.
\end{itemize}
Let us denote the input points by $\bx_i \in  \mathbb{R}^D, i=1, \dots n$, and the $D \times n$ martrix by $\bX=[\bx_1 | \dots | \bx_n]$ whose columns are $\bx_i$. We also denote the corresponding (observed or sampled) input data vector by $\bff=[f(\bx_1), \dots, f(\bx_n)]^T$ and the resulting $n \times n$ covariance matrix by $[\mathbf{K}]_{ij}\equiv K(\mathbf{x}_{i},\mathbf{x}_{j})$. Given the observations $\bff$ on $\bX$, suppose we want to make a probabilistic prediction on $f_* = f(\bx_*)$ at a new point $\bx_*$. Then, GP regression furnishes a probabilistic prediction for the new $\bx_*$ by drawing $f_*$ from the posterior distribution $p(f_* |  \bff, \bX, \bx_*)$.  Using Bayes' Theorem, the conditional property applied to the joint Gaussian prior distribution on $\bff, \bX,$ and $\bx_*$ yields the GP posterior distribution $p(f_* |  \bff, \bX, \bx_*)$ \cite{rasmussen2005,reyes2018new} described by

%%%
\begin{itemize}
\item a newly {\it{updated posterior mean function}}
${\tilde {f}_{*}}\equiv
\bar{f}(\mathbf{x}_{*})+\mathbf{k}_{*}^{T} \mathbf{K}^{-1} \cdot \left(\mathbf{f}-\bar{\mathbf{f}}\right)$, and

%%%
\item a newly {\it{updated posterior covariance}}
$U^{2} \equiv k_{**}-\mathbf{k}_{*}^{T} \mathbf{K}^{-1} \cdot \mathbf{k}_{*}$,
\end{itemize}
%%%
%
where $[\bfk_{*}]_i = K(\bfx_*,\bfx_i)$. Following our previous studies, we regard the updated posterior mean function to be GP's high-order approximation to any agnostic function evaluations at any given point $\bx_*$. Often, the prior mean function is set to zero, $\bar{f}(\bfx) =\bar{\bff}= 0$, to lessen the associated computational footprints, simplifying the expression of the posterior mean function further to
\beq
{\tilde {f}_{*}} = \bz_*^T \cdot \mathbf{f}, \;\;\ \mbox{where} \;\; \bz_*^T = \mathbf{k}_{*}^{T} \mathbf{K}^{-1}.
\label{eq:PostMean}
\eeq
The grid- and kernel-dependent vector $\bz_*^T= \bk_*^T \bK^{-1}$ is called the prediction vector ~\cite{reyes2018new,reyes2019variable}, independent of the data $\bff$, and hence can be calculated, stored, and reused as long as the grid (or the kernel) remains unchanged. If we further restrict ourselves to a `stationary' kernel on a uniform grid (see more discussions on stationary kernels in \cref{sec:Kernel}), only one of these vectors, $\bz_*^T$, needs to be stored for the entire grid. For any given data, $\mathbf{f} \in \mathbb{R}^D$ (e.g., $2r_{gp}+1$ density values $\mathbf{f} = [\rho_{i-r_{gp}}, \dots, \rho_i, \dots, \rho_{i+r_{gp}}]^T$ on a one-dimensional stencil of size $2r_{gp}+1$ with $D=2r_{gp}+1$), the cost of calculating \cref{eq:PostMean} is simply $\mathcal{O}(2D+1)$ floating point operations during the rest of a simulation. Here, the integer-valued $r_{gp}$ is called a GP stencil radius, which is a key factor that determines the order of accuracy of GP solutions.
The posterior covariance conveys an uncertainty in the prediction given by the posterior mean. As such, it can serve as an indicator of how confident the GP prediction is in GP modeling applications. However, it is of limited interest to the purpose of the current paper and its role will be overlooked. See also an extra discussion on the posterior covariance in \ref{apdx:variance}.

%%%%%%%%%%%%%%%%%%%%%%%%%%%%%%% 
%       Section 2.2
%%%%%%%%%%%%%%%%%%%%%%%%%%%%%%%
\subsection{Extending GP's Posterior Prediction to Linear Operations}
To further extend GP regression to broader applications, we first note that symmetric positive definite GP kernels are closed under additions, multiplications, vertical rescaling, convolutions, and linear operations \cite{rasmussen2005,bourgeois2022gp}.  The underlying GP prediction models, such as \cref{eq:PostMean}, can be transformed into another by such operations while preserving the probabilistic GP predictions embedded in the underlying posterior distribution. For the purpose of this paper, we mainly focus on producing new kernels under various linear operators.

Before proceeding further, let us first review the invariance property of GP under linear operators, denoted by $\mathcal{L}$, over the function space GP has generated.
It can be shown that
\beq\label{eq:GP_under_linear}
\mathcal{L}(f) \sim \mathcal{GP}
\bigg(
\mathcal{L} \Big(  m   (\bx ) \Big),
\mathcal{L} \Big( K (\bx, \bx') \Big)
\bigg)
\;\;
\mbox{if} \;\;
f \sim \mathcal{GP}\Big(m(\bx),K(\bx,\bx')\Big),
\eeq 
where 
\beq\label{eq:GP_under_linear2}
\mathcal{L} \Big( K (\bx, \bx')  \Big)  = 
\mathcal{L}_{\bx}  \Big(\mathcal{L}_{\bx'} \Big(K ( \cdot, \bx') \Big)\Big) = 
\mathcal{L}_{\bx'}  \Big(\mathcal{L}_{\bx} \Big(K ( \bx, \cdot) \Big)\Big),
\eeq
depending on whether $\mathcal{L}$ acts on \textit{either $\bx$ or $\bx'$, or both.}
The proof is straightforward and is in  \ref{apdx:GP_closedLinear}.

Using this linearly invariant property, we can generalize the basic GP prediction at $\bx_*$ in \cref{eq:PostMean} to various other applications. A general prediction form can be put into the following form,
\beq\label{eq:GP_general}
p_{*}=\mathcal{L}_2 \left( \mathbf{k}_{*}^{T} \right) \mathcal{L}_1 \left( \mathbf{K}^{-1} \right) \cdot \mathbf{f},
\eeq
where $\mathcal{L}_1$ and $\mathcal{L}_2$ are individual linear operators.
Simplifying, we can write
\beq \label{eq:GP_general_simplified}
p_*= \btt_*^T \bfC^{-1}
\cdot \mathbf{f},
\eeq
where 
$\bfC = \mathcal{L}_1 \Big( K (\bx, \bx')  \Big)$, $\btt_* = \mathcal{L}_2 \left( {K} (\bx_*, \bx') \right)$,  $\mathcal{L}_1$ and $\mathcal{L}_2$ are linear operators applied to their respective kernel matrices, and $p_*$ is our prediction quantity at $\bx=\bx_*$. A trivial case is when both $\mathcal{L}_1$ and $\mathcal{L}_2$ are the point-evaluation functional, e.g., $\mathcal{L}_1\Big( g(\bx) \Big) = \mathcal{L}_2\Big( g(\bx) \Big) = g(\bx)$ for any function $g$, in which case we recover the basic form in \cref{eq:PostMean}. 

A non-trivial example is readily found in our previous studies \cite{reyes2018new,bourgeois2022gp}, where GP's posterior mean function has been extended to a high-order FV reconstruction method. In this case, the input $\mathbf{f}=[\overline{q}_{1}, \dots, \overline{q}_{N} ]^T$ is a one-dimensional vector of length $N$ flattened to contain cell volume-averaged quantities on each cell center at $\bx_h$ (e.g., $\bx_h=(x_{i_h},y_{j_h})$ in 2D Cartesian),
\beq
\overline{q}_{h} 
= \frac{1}{\Delta\mathcal{V}(\bx_h)}\int_{\mathcal{V}(\bx_h)} q(\bx) d\mathcal{V}(\bx),
\eeq
sampled from a local stencil in $D=1,2,3$ spatial dimensions. The output $p_*$ is a pointwise high-order reconstructed Riemann state value.  Here, ${\Delta\mathcal{V}(\bx_h)}$ is the cell volume of the $h$th cell, ${\mathcal{V}(\bx_h)}$, e.g., ${\mathcal{V}(\bx_h)} = I_{h} \times J_{h} = [x_{i_h-1/2}, x_{i_h+1/2}] \times [y_{j_h-1/2}, y_{j_h+1/2}] $ in a 2D Cartesian geometry. 
To account for the input-output relationship, the operator $\mathcal{L}_1$ for $K(\bx,\bx')$ is chosen as a cell volume integral operator over $\bx$ and $\bx'$,
\beq
\mathcal{L}_1 (\cdot) = \frac{1}{\Delta\mathcal{V}(\bx_h)\Delta\mathcal{V}(\bx'_k)}
\int_{\mathcal{V}(\bx_h)} \int_{\mathcal{V}(\bx'_k)}(\cdot)\;\;  d\mathcal{V}(\bx) d\mathcal{V}(\bx'),
\eeq
while $\mathcal{L}_2$ for $\bk_*=K(\bx_*,\bx')$ with a fixed $\bx=\bx_*$ is a volume integral operator over $\bx'$ only,
\beq
\mathcal{L}_2 (\cdot) = \frac{1}{\Delta\mathcal{V}(\bx'_k)}
\int_{\mathcal{V}(\bx'_k)} (\cdot)\;\;  d\mathcal{V}(\bx').
\eeq
Applying $\mathcal{L}_1$ and $\mathcal{L}_2$ to the corresponding kernels produces a GP reconstruction in FVM with the following integrated kernels, $\bfC$ and $\btt_*$, each of which corresponds to their respective basic kernel, $\bfK$ and $\bk_*$.
In componentwise form, they are
\begin{align}
[\bfC]_{hk} &= 
\frac{1}{\Delta\mathcal{V}(\bx_h) \Delta\mathcal{V}(\bx'_k)} 
\int_{\mathcal{V}(\bx_h)} \int_{\mathcal{V}(\bx'_k)}
K(\bx, \bx') 
d\mathcal{V}(\bx) d\mathcal{V}(\bx'), \label{eq:kernel_C}\\
[\btt_{*}]_k &= 
\frac{1}{\Delta\mathcal{V}(\bx'_k)} 
\int_{\mathcal{V}(\bx'_k)}
K(\bx_*,\bx') 
d\mathcal{V}(\bx'). \label{eq:kernel_t}
\end{align}
It is noteworthy to mention the input-output relationship here. The first operator $\mathcal{L}_1$ establishes the covariance relationship of the \textit{input} data at $\bx$ and $\by$, respecting a given data type (e.g., $\bfC$ for volume-averaged and $\bfK$ for pointwise values). The second operator $\mathcal{L}_2$ then takes on the computed covariance and channels the correlations to a scalar \textit{output} in a form we want to obtain (e.g., pointwise Riemann states). This is the main mechanism for how GP can flexibly achieve different types of linear operations on various data types, combine them, and compute an anticipated quantity at $\bx_*$, where the location $\bx_*$ itself is arbitrary, to begin with.

%%%%%%%%%%%%%%%%%%%%%%%%%%%%%%%
%       Section 3
%%%%%%%%%%%%%%%%%%%%%%%%%%%%%%%
\section{Choosing a Kernel Function and GP Stencils} 
In this section, we briefly discuss our options for choosing a GP kernel function and GP local stencil configurations.

%%%%%%%%%%%%%%%%%%%%%%%%%%%%%%%
%       Section 3.1
%%%%%%%%%%%%%%%%%%%%%%%%%%%%%%%
\subsection{Squared Exponential (SE) Kernel}
\label{sec:Kernel}
The kernel calculation of the GP modeling begins with choosing an appropriate GP kernel function, $K$. For our study, we select the squared exponential (SE) covariance kernel,
\begin{equation}
    K(\vec{x}, \vec{x}^{\prime})=K_{\mathrm{SE}}(\vec{x}, \vec{x}^{\prime})=\Sigma^{2} \exp
    \left[-  \frac{ \norm{\vec{x} - \vec{x}^{\prime}}^2}{ 2\ell^2 }  \right]=
    \Sigma^{2} \exp \left[-  \frac{ r^2}{ 2\ell^2 }  \right],
    \label{eq:se}
\end{equation}
which is one of the most widely used GP kernel functions \cite{rasmussen2005,bishop2007pattern} and has been investigated extensively and exclusively in our previous high-order GP studies \cite{lee2017new,reyes2018new,reyes2019variable,reeves2020application,bourgeois2022gp}.
We follow our previous approach to set $\Sigma^{2} = 1$ for simplicity, which does not affect the posterior mean function. The second hyperparameter $\ell$ determines the correlation length scale the GP model prefers. It controls how heavily weighted nearby points are compared to those that are further away in space. There is some flexibility in choosing $\ell$ while an optimal choice is available by minimizing the negative log-likelihood function with respect to $\ell$ \cite{rasmussen2005}. In general, a larger $\ell$ is preferable, but such a choice comes at the cost of making the $\textbf{K}$ a stiffer matrix and harder to invert.

Our choice of the SE kernel is supported by three prime attractive properties it affords. First, the Gaussian function delivers a $C^{\infty}$ prior function space, which guarantees that its data fitting accuracy by those sampled admissible $C^{\infty}$ functions is convergent linearly with its stencil size. The built-in regularity also provides the SE function with analytic forms of derivatives and integrations, which makes SE much more accessible to various applications.
Second, if we assume a constancy of the two hyperparameters, $\Sigma^2$ and $\ell$, the kernel becomes a function of the Euclidean distance between $\bx$ and $\bx'$ only, hence making it a \textit{stationary} kernel. It is also \textit{data-independent} as described in \cref{sec:GP_prior_posterior}. These two points together greatly simplify the GP prediction in \cref{eq:PostMean} to a simple dot product of $\mathcal{O}(2D+1)$ since $\bz_*^T$ can be pre-computed in an initial step. 
Third, the SE kernel facilitates dimensional factorization, i.e.,
\beq
\exp  \left[- \frac{r^2}{2\ell^2} \right] = \prod_{d} \exp  \left[- \frac{r_d^2}{2\ell^2} \right],
\label{eq:se_arbdim}
\eeq
where the radial distance between $\bx$ and $\bx'$ is $r^2 = \Sigma_d r_d^2$ with $r_d$ in each $d$ dimension.
This factorization property enables a seamless extension from 1D to a higher spatial dimension, making our GP prediction capability readily feasible in multidimensional cases.

Unfortunately, these desirable properties of SE could be severely compromised when $\bx \to \bx'$ or $\ell \to \infty$, which results in a poorly conditioned kernel matrix $\bfK$. In practice, the latter case is less of an issue as we choose $\ell$ a bounded value. Our typical choice for $\ell$ is either a constant that represents a characteristic length scale of a given profile or a scale distance proportional to the grid scale set by $\ell = M \Delta$ with $\Delta = \min_d{\Delta_d}$ and $6 \le M \le 12$. See more discussions on $\ell$ in \cite{reyes2018new,reyes2019variable,reeves2020application,bourgeois2022gp}.
The former case with $\bx \to \bx'$  causes the singularity issue more frequently as this scenario occurs as one resolves the grid scales, $\Delta \to 0$. We have documented several different approaches that address the issue, most recently in \cite{bourgeois2022gp} as well as in \cite{reyes2018new,reyes2019variable,reeves2020application}. The simplest (and easiest) approach is adopted in the current study, which utilizes quadruple precision in compiling only a couple of routines that are relevant to calculate $\bz_*^T$ once and for all.  Unlike the usual doubt, the added computational cost of handling quadruple-precision is a bare minimum and does not impact the overall performance of GP at all. Interested readers are encouraged to refer to the relevant discussions in our work \cite{reyes2018new,reyes2019variable,bourgeois2022gp}.

It is worth noting the different types of GP kernels, particularly those that are non-smooth. Although SE performs well on smooth datasets, its posterior prediction is prone to underperform with oscillations when datasets contain discontinuities that are common in shock-dominant fluid dynamics simulations. There are better-suited non-smooth kernels such as the Mat\'{e}rn class, Wendland, rational quadratic functions, a neural network covariance kernel function, etc. \cite{rasmussen2005,stein2012interpolation,fasshauer2015kernel} for discontinuous datasets. On smooth data, though, these kernels exhibit limited usability because their solution accuracy, modeled by the latent functions with finite-order differentiability, is reduced and lower than that of SE, independent of the input data size $\bff$. This is in contrast to the fact that the accuracy of SE increases linearly (or decreases) with increasing (or decreasing) size of the input data, which can be readily controlled by the GP stencil configuration. Later in \cref{sec:gp_das}, we introduce a new discontinuous kernel that is superior in fitting discontinuous data without oscillation while it converges at second-order on smooth data. We discuss GP stencils in the next section. 

%%%%%%%%%%%%%%%%%%%%%%%%%%%%%%%
%       Section 3.2
%%%%%%%%%%%%%%%%%%%%%%%%%%%%%%%
\subsection{1D GP Stencils}
\label{sec:1DStencil}

We take a vector of size $(2r_{gp}+1)$ for our 1D GP stencils, containing contiguous cells from $x_{i-r_{gp}}$ to $x_{i+r_{gp}}$ centered at $x_i$. With this stencil configuration, a native point-to-point GP interpolation that takes an input $\mathbf{f}=[q_{i-r_{gp}}, \ldots, q_{i+r_{gp}}]^{T}$ for a function $q$ with $q_j=q(x_j)$ to produce a function interpolation $q(x_*)=q_*$ with $x_* \in [x_{i-r_{gp}}, x_{i+r_{gp}}]$ becomes a $(2r_{gp}+1)$-th accurate interpolation scheme following the stencil size linearly \cite{reyes2018new,reyes2019variable,bourgeois2022gp}. Our default choice of $x_*$ in 1D is $x_{i\pm 1/2}$, the $i$-th cell interface locations, where various numerical methods (e.g., finite difference, finite volume, discontinuous Galerkin, etc.) calculate high-order numerical quantities such as Riemann states and numerical fluxes.

%%%%%%%%%%%%%%%%%%%%%%%%%%%%%%%
%       Fig. 1
%%%%%%%%%%%%%%%%%%%%%%%%%%%%%%%
\begin{figure}[H]
  \centering
  \includegraphics[width=0.6\textwidth]{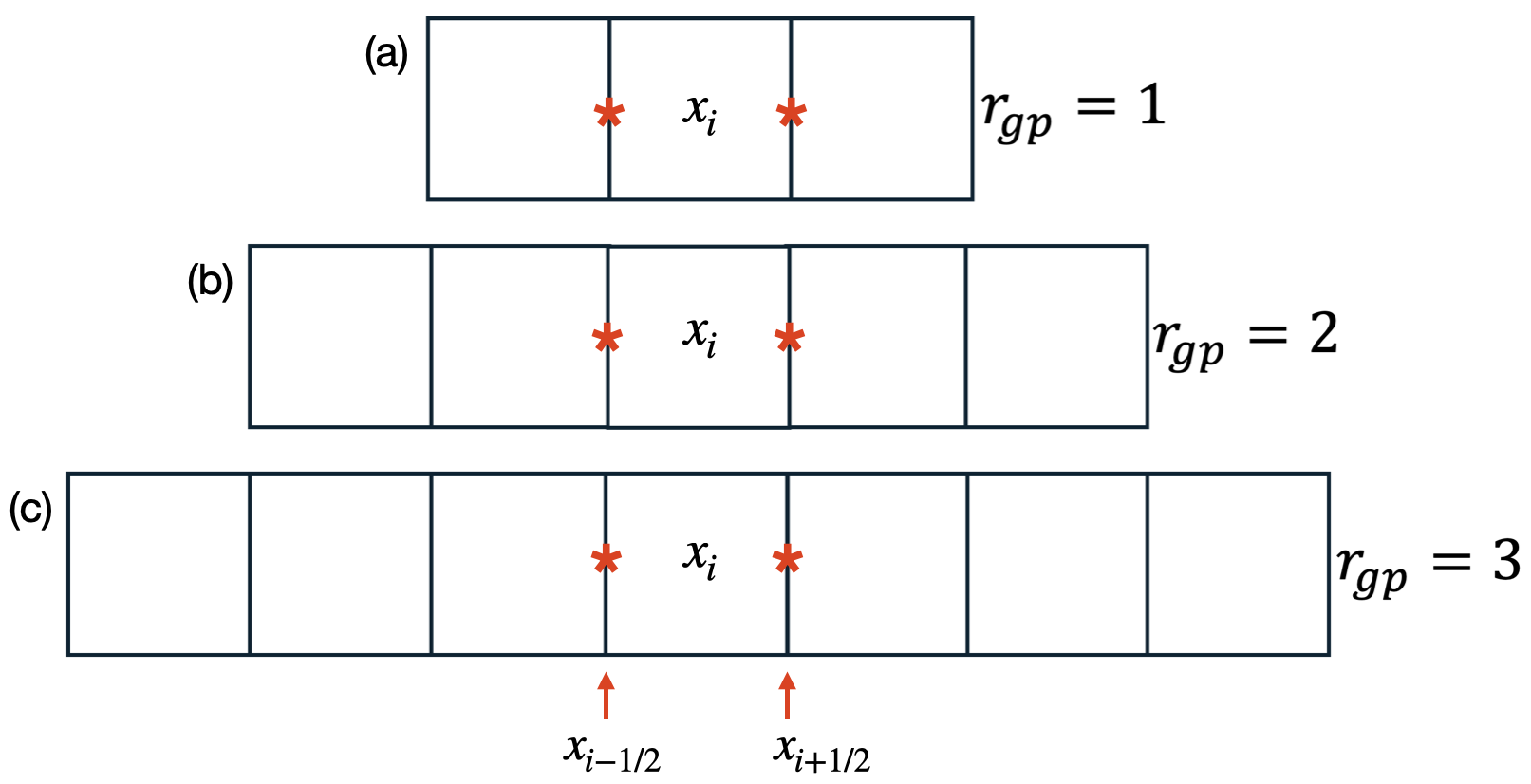}  
  \caption{1D GP stencil. The red ``*'' marks indicate the $i$-th cell interface locations we choose for $x_*$ in 1D. The accuracy of the native GP point interpolations at $x_*$ is $(2r_{gp}+1)$, showing (a) the 3rd-order stencil, (b) the 5th-order stencil, and (c) the 7th-order stencil, where $r_{gp}$ denotes the respective integer-valued GP stencil radius from the cell center $x_i$. A higher-order GP stencil can be easily configured with $r_{gp} > 3$.}
  \label{fig:gp_1d_stencil}
\end{figure}

%%%%%%%%%%%%%%%%%%%%%%%%%%%%%%%
%       Section 3.3
%%%%%%%%%%%%%%%%%%%%%%%%%%%%%%%
\subsection{2D GP Stencils}
\label{sec:2DStencil}
\begin{figure}[h!]
  \centering
  \includegraphics[width=0.7\textwidth]{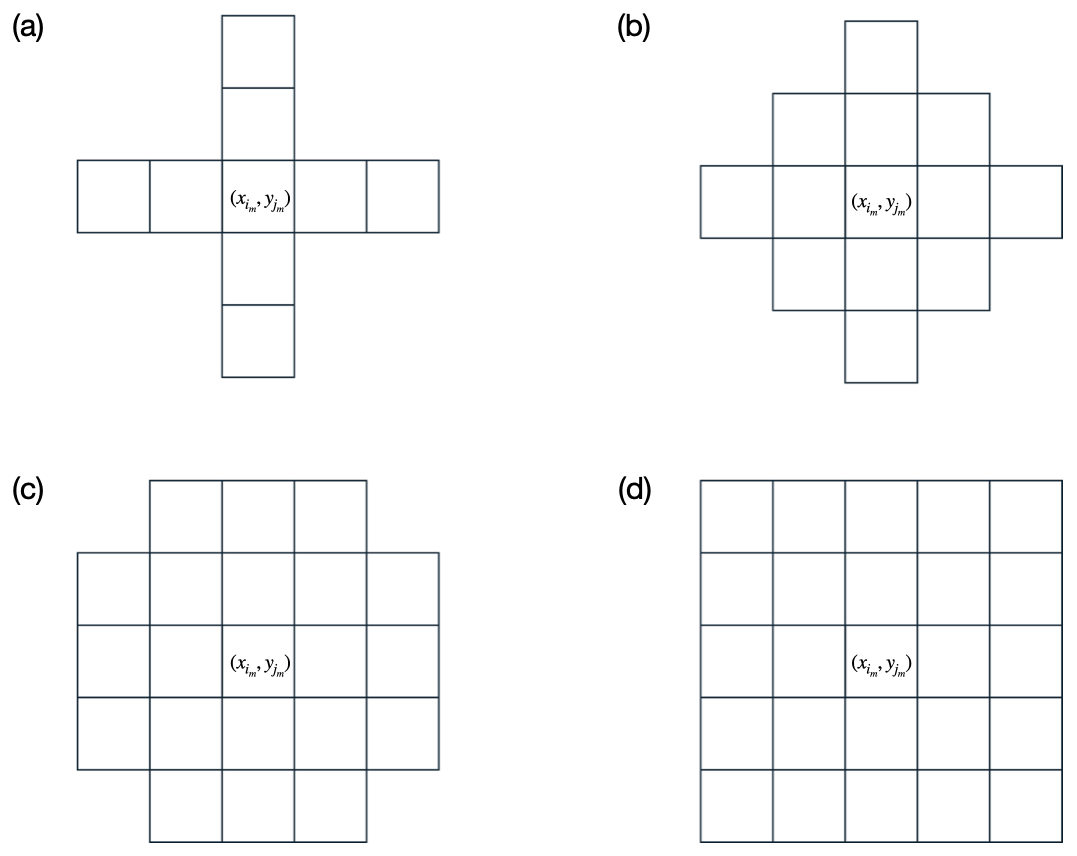}  
  \caption{2D GP stencils of $r_{gp}=2$. The central cell $\bx_m = (x_{i_m},y_{j_m})$ is the cell where GP predictions are computed at the surrounding interface locations, $(x_{i_m \pm 1/2},y_{j_m})$ or $(x_{i_m},y_{j_m \pm 1/2})$, or corner locations, $(x_{i_m \pm 1/2},y_{j_m\pm 1/2})$.}
  \label{fig:gp_2d_stencil}
\end{figure}

It is quite flexible to extend 1D GP stencils to a higher spatial dimension. This is yet another feature that makes GP regression genuinely multidimensional and advantageous in handling multidimensional stencil data over the existing polynomial counterparts. For instance, in \cref{fig:gp_2d_stencil}, we display four different GP stencil configurations in 2D with $r_{gp}=2$ centered at $\bx_m=(x_{i_m}, y_{j_m})$, including (a) the cross stencil, (b) the diamond stencil, (c) the blocky-diamond stencil, and (d) the square stencil. 
%
%%%%%%%%%%%%%%%%%%%%%%%%%%%%%%%
%       Fig. 2
%%%%%%%%%%%%%%%%%%%%%%%%%%%%%%%

To decide on an optimal 2D stencil, we examine the so-called \textit{effective order of convergence} (or EOC) rates for various GP prediction operations on each configuration (more details to follow in \cref{sec:stencil_selection}) and choose the most compact stencil configuration that meets the expectedlinear convergence rate, e.g.,  $(2r_{gp}+1)$ for the baseline point-to-point interpolation. See more applications in \cref{sec:stencil_selection}.
On the cross stencil in (a), EOC falls short of delivering the anticipated convergence rate on calculations involving any off-axis information simply because it includes the axis-aligned cells only. The diamond stencil in (b) adds four extra corner cells to the cross configuration in (a), and as a result, it demonstrates a more improved accuracy than (a) when multidimensional calculations are considered in particular. It was indeed chosen as the default configuration for all 2D FV simulations in \cite{bourgeois2022gp}. Yet, we found it delivers a reduced convergence rate in some applications considered in the current study, and we opted it out as a default choice. 
The two configurations in (c) and (d) successfully meet the anticipated convergence rates %$\mathcal{O}(2r_{gp}+1)$ 
on various applications. While their performances on convergence rates are equivalent, the ``blocky diamond'' configuration (c) has a smaller stencil size than the full square stencil in (d). This makes us choose the blocky diamond shape in (c) for our default GP stencil configuration. We also remark that 3D stencil configurations can be decided in a similar way, although we mainly focus on 2D calculations in the current study. 

%%%%%%%%%%%%%%%%%%%%%%%%%%%%%%%
%       Section 4
%%%%%%%%%%%%%%%%%%%%%%%%%%%%%%%
\section{Stencil Selection and Convergence Analysis}
\label{sec:stencil_selection}
It remains to discuss how to examine GP convergence rates on each configuration to select an optimal GP stencil for each $r_{gp}$. Selecting an appropriate stencil is crucial in our GP methods as it directly influences the accuracy, efficiency, and stability of predictions. 
One conventional approach would be to perform grid convergence studies on a set of grid resolutions for each grid configuration and select one that is optimal in the sense we described in \cref{sec:2DStencil}. This strategy is absolutely necessary to validate the model's accuracy in general, particularly when the model's convergence rate is theoretically deduced and can be determined by design as in various polynomial approximations that are based on Taylor series expansions.
However, as a kernel-based non-polynomial method, GP lacks such theoretical design principles, which leaves us to run numerous grid convergence tests to choose an optimal stencil configuration, making a stencil selection itself a computationally daunting task. To ameliorate the lack of such theoretical tools for GP, we developed a systematic approach to illustrate (and predict) GP's \textit{effective} convergence rates without executing expansive convergence tests. We introduce this methodology in this section.

A key question we address is how to determine the optimal stencil and predict its \textit{effective} rate of convergence. The objective is to identify a stencil that achieves the highest accuracy while minimizing computational cost. To answer this question, we analyze the effective accuracy of different stencils based on the structure of the kernel weights and their interaction with the surrounding grid data points. In conventional numerical methods, the order of accuracy is typically determined by a Taylor series expansion of the approximation. However, unlike conventional methods explicitly designed to cancel lower-order error terms, GP methods approximate function values at $\bx_*$ through kernel-weighted interpolation. This results in an \textit{effective} order of accuracy, which may differ from the \textit{expected} or \textit{anticipated} order of accuracy due to the leading error term in conventional polynomial-based finite difference approximations.

%%%%%%%%%%%%%%%%%%%%%%%%%%%%%%%
%       Section 4.1
%%%%%%%%%%%%%%%%%%%%%%%%%%%%%%%
\subsection{Taylor Expansion in GP Predictions to Analyze GP's Effective Rate of Convergence}
\label{sec:gp_order_analysis}

To systematically analyze GP's effective order of accuracy, we leverage a Taylor series expansion approach in GP-produced approximations. This method is inspired by classical error analysis in finite difference methods \cite{doi:10.1137/1.9780898717839} and extends the framework proposed in \cite{mishra2023orderaccuracyfinitedifference} to non-standard numerical methods, such as GP-based schemes, where perfect error term cancellation is not guaranteed.

First, we note that GP predictions involve kernel-weighted interpolation. This means that once the kernel calculations have been computed, the prediction simplifies to a weighted sum of the data,
\begin{equation}
    f_{\text{GP}}(x_*) = \sum_{i} w_i f(x_i),
    \label{eq:GP_weighted_sum}
\end{equation}
where the weight vector \( \mathbf{w} \) with components $w_i$ is given by the prediction vector as defined in \cref{eq:PostMean}
or more generally in \cref{eq:GP_general_simplified},
\begin{equation}
    \mathbf{w} = \btt_*^T \bfC^{-1}.
\end{equation}
For 1D GP predictions on the GP stencil of size \( (2r_{gp}+1)\), we Taylor expand each \( f(x_i) \) at the point \( x=x_* \) to recast \cref{eq:GP_weighted_sum} into a Taylor-expanded GP prediction,
\begin{equation}
    f_{\text{GP}}(x_*) = \sum_{i=1}^{2r_{gp}+1} w_i \left( \sum_{\alpha=0}^{k} \frac{f^{(\alpha)}(x_*)}{\alpha!} (x_i - x_*)^\alpha \right) + \mathcal{O}(h^{\alpha+1}), 
\end{equation}
where $h=x_i - x_*$. This formulation explicitly reveals how the kernel weights influence the coefficients of the Taylor expansion terms and, ultimately, determine the accuracy of the prediction.

This strategy is readily extended to GP predictions in 2D and 3D by using the Taylor expansion for a function $f:\mathbb{R}^d \to \mathbb{R}$, $d=2$ or $3$,
\begin{equation}
    f(\bx_i) = \sum_{|\boldsymbol\alpha| =0}^{k} \frac{\partial^{|\boldsymbol\alpha|} f(\bx_*)}{\partial\bx^{\boldsymbol\alpha}} \frac{\bh^{\boldsymbol\alpha}}{\boldsymbol\alpha !} + \mathcal{O}(\mathbf{\bh}^{\tilde{\boldsymbol\alpha}})
    = \sum_{|\boldsymbol\alpha| =0}^{k} \frac{\partial^{|\boldsymbol\alpha|} f(\bx_*)}{\partial x_1^{\alpha_1} \cdots \partial x_n^{\alpha_n}} \frac{h_1^{\alpha_1} \cdots h_n^{\alpha_n}}{\boldsymbol\alpha !} + \mathcal{O}(\mathbf{\bh}^{\tilde{\boldsymbol\alpha}}),
\end{equation}
where $\bh = \bx_i - \bx_*$. Following the standard convention, $\boldsymbol\alpha = (\alpha_1, \dots, \alpha_d)\in \mathbb{N}^d_0$ is an $d$-tuple multi-index with $|\boldsymbol\alpha|=\sum_{i=1}^d \alpha_i$, where $|\tilde{\boldsymbol\alpha}|=\sum_{i=1}^d \alpha_i = k+1$.

%%%%%%%%%%%%%%%%%%%%%%%%%%%%%%%
%       Section 4.2
%%%%%%%%%%%%%%%%%%%%%%%%%%%%%%%
\subsection{Effective Order of Convergence (EOC)}\label{sec:gp_eoc}
In traditional numerical methods, the formal order of accuracy is determined by the highest nonzero term in the Taylor series. However, in GP-based methods, lower-order terms may persist with small but nonzero coefficients, affecting the actual observed accuracy.
To quantify the overall contribution of different terms, we introduce the effective order of convergence (EOC) defined as,
\begin{equation}
  {\text{EOC}} = \frac{\sum_{h=1}^{k} h |c_h|}{\sum_{h=1}^{k} |c_h|},
  \label{eq:eoc}
\end{equation}
where \( k \) is the number of the terms in the Taylor expansion, \( c_h = \frac{\partial^{|\boldsymbol\alpha|} f(\bx_*)}{\partial\bx^{\boldsymbol\alpha}} \frac{1}{\boldsymbol\alpha !}\) with $|\boldsymbol\alpha|=h$ are the coefficients of the expansion terms, and 
EOC represents the effective observed order of accuracy. This metric provides a holistic view of GP accuracy, accounting for all error contributions rather than only the dominant term. We set $k=12$ for the current study.

%%%%%%%%%%%%%%%%%%%%%%%%%%%%%%%
%       Section 4.3
%%%%%%%%%%%%%%%%%%%%%%%%%%%%%%%
\subsection{Selecting the Optimal GP Stencil based on EOC}
\label{sec:stencil_selection_eoc}
Using the effective order of convergence in \cref{eq:eoc}, we now examine the numerical convergence order of different 2D stencils to determine which configuration provides the most effective accuracy. This selection process is unnecessary for 1D GP stencils as their configuration is uniquely determined by the GP radius $r_{gp}$, and 1D GP predictions follow linear convergence rates as expected.

To demonstrate, we computed EOCs on four different GP stencil configurations, (a) -- (d) in \cref{fig:gp_2d_stencil}, for six different GP operations (see \cref{tab:stencil_convergences}) where input vector $\bff$ contains cell-centered pointwise values $[\bff]_i = f_i$ sampled from each GP stencil. We display the respective results in \cref{tab:stencil_convergences}.
Also shown as reference in the third column of \cref{tab:stencil_convergences} are the so-called `anticipated target orders of accuracy' we would normally expect in, e.g., 1D conventional finite difference approximations, assuming that the baseline interpolation $\bff \to f(\bx_*)$ follows the $(2r_{gp}+1)$th convergence rate to begin with. In this sense, we expect that the anticipated order of accuracy should decrease by one from the baseline $(2r_{gp}+1)$ rate as each differential approximation is added to the approximating operation. In all test cases, we set the GP radius by $r_{gp}=2$, and the baseline interpolation is expected to be fifth-order accurate. Approximations to partial derivatives are considered at three different locations of $\bx_*$, including the central cell location $(x_{i_m}, y_{j_m})$, the $x$ interface location $(x_{{i_m}\pm 1/2}, y_{j_m})$, and the cell corner location $(x_{{i_m}\pm 1/2}, y_{{j_m}\pm 1/2})$. We omit the $y$ interface calculations since they are symmetric calculations of the $x$ interface results, so we don't repeat them.

            \begin{table}[ht!]
            \caption{Anticipated target orders (third column) and GP's effective order of convergence (EOC in \cref{eq:eoc}) for one data interpolation and five different derivative approximations at $\bx_*$ (fourth -- sixth columns) on four different GP stencils with $r_{gp}=2$. Input data are the data vector $\bff$ whose components $f_i$ are cell-centered pointwise values sampled from each stencil. EOCs are calculated at (i) the central cell $(x_{i_m},y_{j_m})$ in the fourth column, (ii) the $x$ interfaces  $(x_{{i_m}\pm 1/2}, y_{j_m})$ of the central cell in the fifth column, and (iii) the cell corners $(x_{{i_m}\pm 1/2}, y_{{j_m}\pm 1/2})$ of the central cell in the sixth column.}
            \centering
            \scriptsize
            %\footnotesize
            \begin{tabular}{|c|l|c|c|c|c|}
                \hline
                {GP stencil} & {Operation} & Anticipated Order & $\text{EOC at } \bx_* = (x_{i_m}, y_{j_m})$ & $\text{EOC at } \bx_* = (x_{{i_m}\pm 1/2}, y_{j_m})$ & $\text{EOC at } \bx_* = (x_{{i_m}\pm 1/2}, y_{{j_m}\pm 1/2})$\\
                \hline\hline
                                               & $\bff \to f(\bx_*)$               &5& \text{Exact} & 5.2174 & 3.0330 \\
                                               & $\bff \to \partial_x f(\bx_*)$    &4& 3.8093 & 3.6448 & 1.3237 \\
                {Cross}                        & $\bff \to \partial_y f(\bx_*)$    &4& 3.8093 & 1.6109 & 1.3237 \\
                {(\cref{fig:gp_2d_stencil}(a))}& $\bff \to \partial_{xx} f(\bx_*)$ &3& 3.5950 & 3.2457 & 2.1973 \\
                                               & $\bff \to \partial_{xy} f(\bx_*)$&3& 0.0000 & 1.9361 & 1.3040 \\
                                               & $\bff \to \partial_{yy} f(\bx_*)$ &3& 3.5950 & 1.3944 & 2.1973 \\        
                \hline\hline
                                                & $\bff \to f(\bx_*)$               &5& \text{Exact} & 5.2174 & 5.1315 \\
                                                & $\bff \to \partial_x f(\bx_*)$    &4& 3.8093 & 3.6448 & 3.4121 \\
                {Diamond}                       & $\bff \to \partial_y f(\bx_*)$    &4& 3.8093 & 3.6555 & 3.4121 \\
                {(\cref{fig:gp_2d_stencil}(b))} & $\bff \to \partial_{xx} f(\bx_*)$ &3& 3.5950 & 3.2457 & 3.2984 \\
                                                & $\bff \to \partial_{xy} f(\bx_*)$ &3& 2.2689 & 2.1356 & 2.0204 \\
                                                & $\bff \to \partial_{yy} f(\bx_*)$ &3& 3.5950 & 3.4222 & 3.2984 \\
                \hline \hline
                                                & $\bff \to f(\bx_*)$               &5& \text{Exact} & 5.2171 & 5.7126 \\
                                                & $\bff \to \partial_x f(\bx_*)$    &4& 3.8093 & 3.6448 & 5.0371 \\
                {Blocky Diamond}                & $\bff \to \partial_y f(\bx_*)$    &4& 3.8093 & 4.7600 & 5.0371 \\
                {(\cref{fig:gp_2d_stencil}(c))} & $\bff \to \partial_{xx} f(\bx_*)$ &3& 3.5950 & 3.2457 & 4.0972 \\
                                                & $\bff \to \partial_{xy} f(\bx_*)$ &3& 4.0719 & 3.8882 & 3.7237 \\
                                                & $\bff \to \partial_{yy} f(\bx_*)$ &3& 3.5950 & 4.8469 & 4.0972 \\
                \hline\hline
                                                & $\bff \to f(\bx_*)$               &5& \text{Exact} & 4.9229 & 4.9844 \\
                                                & $\bff \to \partial_x f(\bx_*)$    &4& 3.8121 & 3.7314 & 4.9225 \\
                {Square}                        & $\bff \to \partial_y f(\bx_*)$    &4& 3.8090 & 4.3776 & 4.9481 \\
                {(\cref{fig:gp_2d_stencil}(d))} & $\bff \to \partial_{xx} f(\bx_*)$ &3& 3.6626 & 3.2551 & 3.5499 \\
                                                & $\bff \to \partial_{xy} f(\bx_*)$ &3& 3.9009 & 3.8126 & 3.7697 \\
                                                & $\bff \to \partial_{yy} f(\bx_*)$ &3& 3.6375 & 4.6015 & 3.5523 \\
                \hline\hline
            \end{tabular}
            \label{tab:stencil_convergences}
        \end{table}

We aim to choose an optimal stencil configuration that can deliver its EOCs as much as close to the anticipated order of convergence. First, we see that the EOCs on the cross stencil achieve the fifth-order accuracy for the baseline point-to-point interpolation case at the $x$ interface. However, the stencil delivers only the third-order convergence at the corner location. This is expected since the stencil lacks points along the diagonal direction, which carries important information in diagonally-oriented calculations. A similar deficiency is also found in the cross derivative calculation $\partial_{xy}f(\bx_*)$ that only achieves an EOC $\sim 1.3$, lower than the other grid-aligned counterparts, $\partial_{xx}$ or $\partial_{yy}$. Nevertheless, EOCs fall short of the anticipated orders on the remaining first and second derivative test cases at the cell corners, which makes us conclude that the most compact  stencil is suboptimal and inappropriate.
Second, the results on the diamond stencil prove to be better than those on the cross case in derivative tests, although they fail to show the anticipated fourth- and third-order rates in some cases, especially on the cross derivative case.  Comparing this EOC $\sim 2.02$ with the cross stencil EOC  $\sim 1.3$, adding the extra four corner cells to the cross configuration generally helps improve the diagonal calculation, except that it is insufficient enough to meet the third-order anticipated accuracy.

In comparison, the EOCs on the blocky diamond and square stencils are shown to produce orders that match the anticipated orders. In fact, we see that the EOCs are higher than the anticipated accuracy in most cases, convincing the validity of the stencil configurations. From this analysis, we select the blocky diamond as our default configuration since it delivers qualitatively equivalent results to the ones on the square stencil while being a bit more compact. 

There are a couple of notes we want to discuss. First, on all stencils, the baseline interpolation $\bff \to f(\bx_*)$ at the central cell $(x_{i_m},y_{j_m})$ is exact  due to the property of GP to exactly recover data given to its prior distribution.  Second, it is worth pointing out  one outlier case  $\bff \to \partial_{xy} f(\bx_*)$ on the cross stencil, which returns a convergence of exactly 0 due to the perfect cancellations in the leading term. This can be seen in \cref{eq:se_2d_dxdy} that $(-2x + 2x')$ will always be 0 in the vertical strip of the cross stencil and $(-2y + 2y')$ will always be zero in the horizontal strip, leading to effectively zero contributions from all points on the stencil. 

Lastly, we make observations on the convergence of the derivative terms, which is shown to behave differently from the anticipated order. As discussed, we normally expect a drop in order of accuracy for each derivative taken, so we would naively look for at least  fourth-order convergence for  $\bff \to \partial_{x} f(\bx_*)$ and $\bff \to \partial_{y} f(\bx_*)$. For example, the $x$ interface calculations of the $x$ derivative are equivalent between the diamond and blocky diamond stencils with EOCs $\sim3.6448$, close to the anticipated fourth-order convergence. Although a similar rate is observed in the $y$ derivative EOC on the diamond stencil ($\sim 3.6555$), it is interesting to see a higher EOC $\sim 4.76$ is achieved on the blocky diamond for the same problem. Similar higher orders are also found on the blocky diamond on the other second derivative tests involving the $y$ derivative as well. The $y$ direction improvements on the blocky stencil can be seen by looking at the number of cells along the $y$ direction, above and below the $x$ interface location $x_{i\pm 1/2}$. The diamond stencil has three points, while the blocky diamond has five. This added directional contribution to the blocky configuration results in an EOC that is indeed higher than the anticipated fourth order. This reasoning explains the identical EOCs at the corner location $(x_{{i_m}\pm 1/2}, y_{{j_m}\pm 1/2})$ on the blocky stencil, e.g., EOCs $\sim 5.0371$ between $\bff \to \partial_{x} f(\bx_*)$ and $\bff \to \partial_{y} f(\bx_*)$ and the EOCs $\sim 4.0972$ between $\bff \to \partial_{xx} f(\bx_*)$ and $\bff \to \partial_{yy} f(\bx_*)$ as the participating cell counts are consistently symmetric in $x$ and $y$ directions. The results of the blocky diamond and the square stencils are quite similar, and they both perform well in meeting the anticipated orders. It may seem a bit strange why some of the EOCs on the square stencil are lower than those on the blocky stencil. We suspect that this counter-intuitive behavior could be related to an overfitting issue, where addition of data points beyond what is necessary may overfit the data and degrade GP predictions.

We now draw our conclusion. Our Taylor series-based analysis for GP provides powerful insights into tuning GP's prediction capabilities for a set of linear operators on different stencil configurations. We used this approach to identify an optimal GP stencil configuration across a selected set of GP operations in this section. We emphasize that the results here will be redone and cross-checked on actual grid convergence tests to validate the grid-based GP performance on a broad range of linear operators on our optimal GP stencil. The EOC comparison study illustrates that the blocky diamond in \cref{fig:gp_2d_stencil}(c) performs well across different types of derivative approximations, which makes us select it as our optimal (and default) stencil choice in 2D.

%%%%%%%%%%%%%%%%%%%%%%%%%%%%%%%
%       Section 5
%%%%%%%%%%%%%%%%%%%%%%%%%%%%%%%
\section{List of Applications}
In the rest of the paper, we present numerical results to validate the findings in the previous section further. We begin by providing a list of selected applications based on their popularity in modeling numerical differential equations. Most frequent operations include derivatives and integrals applied to pointwise and volume-averaged data types. We limit our interests to 1D and 2D applications only, although extensions to 3D applications are straightforward.

\begin{table}[h!]
  \centering
  \caption{A list of GP-recipe applications, including the baseline point-evaluation operator denoted by PE, e.g., $\mathcal{L}(g(\bx)) = g(\bx)$ for a function $g$. The input is a $N$-vector $\bg$ whose $h$th component $[\bg]_h = g(\bx_h)$ is shown in the first column. The output in the second column is a scalar function value evaluated at $\bx=\bx_*$. 
  Following \cref{eq:GP_general}, $\mathcal{L}_1$ is applied to $K(\bx,\bx')$ and $\mathcal{L}_2$ is applied to $K(\bx_*,\bx')$, acting on either $\bx$ or $\bx'$, or both. For simplicity, we assume that all calculations are performed using Cartesian geometry. In the first and second columns, we also show three abbreviations, ``pt'',  ``va'', and ``der'', representing pointwise, volume-averaged, and derivative data types in each input and output pair. The notation $I_{h}$ is reserved to denote a 1D cell $[x_{h - 1/2}, x_{h + 1/2}]$ centered at $x_{h}$, and similarly for $J_{k}$ centered at $y_{k}$, as well as for $I_{*}$ centered at $x_*$ and for  $J_{*}$ centered at $y_*$. }  
  \footnotesize
  %\scriptsize
  \begin{tabular}{|c|cc|cc|cc|cc|}%{|l|lc|lc|lc|} %{|c|cc|cc|cc|}
    \hline 
\multirow{2}{*}{1D Input } & \multirow{2}{*}{1D Output} & & \multicolumn{2}{c|}{$x, x' \in \mathbb{R}$}  & \multirow{2}{*}{Section} &\\ \cline{4-5}  & &     & $\mathcal{L}_1(K(x,x'))$  & $\mathcal{L}_2(K(x_*,x'))$  &  &\\ \hline \hline
%%%%%%%%%%%%%%%%%%%% 1D %%%%%%%%%%%%%%%%%%%%%%%%%%%%%%%%%%%%%%%%%
$f(x_h)$  (pt)&$f(x_*)$ (pt)&     &  PE  & PE &    \cref{sec:pt-to-pt_1D}  &   \\    
$f(x_h)$  (pt)&$f^{(m)}(x_*)$ (der)&    &  PE  & $ \frac{d^m}{dx^m} (\cdot)$  &    \cref{sec:pt-to-der_1D}  &   \\   
$f(x_h)$  (pt)& $\frac{1}{\Delta x}\int_{I_{*}} f(x) dx$ (va)&   & PE  & $\frac{1}{\Delta x} \int  (\cdot) \; dx$      & \cref{sec:pt-to-va_1D}  & \\
$\frac{1}{\Delta x}\int_{I_{h}} f(x) dx$ (va)&   $f(x_*)$ (pt)&    & $\frac{1}{\Delta x \Delta x'}\int \int (\cdot) \; dx dx'$ & $\frac{1}{\Delta x'} \int  (\cdot) \; dx'$     & \cref{sec:va-to-pt_1D} &    \\ \hline \hline
%%%%%%%%%%%%%%%%%%%%% 2D %%%%%%%%%%%%%%%%%%%%%%%%%%%%%%%%%%%%%%%%%
\multirow{2}{*}{2D Input } & \multirow{2}{*}{2D Output} & & \multicolumn{2}{c|}{$\bx=(x,y), \bx'=(x',y') \in \mathbb{R}^2$} & \multirow{2}{*}{Section} &\\ \cline{4-5}  & & &  $\mathcal{L}_1(K(\bx,\bx'))$  &$\mathcal{L}_2(K(\bx_*,\bx'))$  &  &\\ \hline \hline
$f(x_{h},y_{k})$  (pt)&$f(x_*,y_*)$ (pt)&  & PE    & PE   &  \cref{sec:pt-to-pt_2D} &     \\   
$f(x_{h},y_{k})$   (pt)& $\partial_y^n \partial_x^{m} f(x_*,y_*)$ (der)&     & PE & $\partial_{y}^n \partial_{x}^{m} (\cdot)$   & \cref{sec:pt-to-der_2D} &     \\
%%% 
$f(x_{h},y_{k})$   (pt)& $\frac{1}{\Delta x\Delta y}\int_{I_{*}} \int_{J_{*}} f(x,y) dxdy$ (va) &     & PE & $\frac{1}{\Delta \bx}\int (\cdot) \; d\bx$    & \cref{sec:pt-to-va_2D} &     \\
$\frac{1}{\Delta x\Delta y}\int_{I_{h}} \int_{J_{k}} f(x,y) dxdy$  (va)&  $f(x_*,y_*) $ (pt)&     & $\frac{1}{\Delta \bx \Delta \bx'}\int \int (\cdot) \; d\bx d\bx'$      & $\frac{1}{\Delta \bx'} \int (\cdot) \; d\bx' $   & \cref{sec:va-to-pt_2D} &     \\
$\frac{1}{\Delta x\Delta y}\int_{I_{h}} \int_{J_{k}} f(x,y) dxdy$  (va)& $ \partial_y^n \partial_x^{m} f(x_*,y_*) $ (der)&  & $\frac{1}{\Delta \bx \Delta \bx'}\int \int (\cdot) \; d\bx d\bx'$  & $\partial_{y}^{n} \partial_{x}^{m} \left(\frac{1}{\Delta \bx'} \int  (\cdot) \; d\bx' \right)$  & \cref{sec:va-to-der_2D}&\\ \hline
%%%%%%%%%%%%%%%%%%%%% End of 2D %%%%%%%%%%%%%%%%%%%%%%%%%%%%%%%%%%%%%%%%%
\end{tabular}
\label{tab:GP_applications}
\end{table}

\cref{tab:GP_applications} summarizes a set of linear operations that are particularly of interest in discretizing partial differential equations. The input $\bg$ is a 1D vector of length $N$ that contains all participating grid data values from a local GP stencil (see \cref{fig:gp_1d_stencil,fig:gp_2d_stencil}), which are concatenated (or flattened) into a 1D vector form if the data values are collected from a multi-dimensional stencil. As discussed in the previous section, our default GP stencil in 2D is the blocky diamond shape in \cref{fig:gp_2d_stencil}(c). 

Common types of input data values appearing in numerical modeling are shown in the first column, showing only a single component. Target outputs are displayed in the second column, which are scalar values that represent linearly transformed function evaluations at the point of interest, $\bx=\bx_*$. For exposition purposes, we consistently choose the variable $\bx$ (over $\bx'$) to assign the evaluation point, $\bx=\bx_*$, although the roles of $\bx$ and $\bx'$ are interchangeable in general.
The $\mathcal{L}_1$ operation transforms the kernel function $K$ linearly to construct the covariance kernel matrix $\bfK$, embedding the two-point correlations  between the input data values at $\bx=\bx_h$ and $\bx'=\bx_k$ into $[\bfK]_{hk}$ in the data type prescribed in the first column. Such correlations are then channeled into the output via $\mathcal{L}_2$ by translating the correlated input data information into the target output at $\bx_*$ in various data types. The operations of the $\mathcal{L}_2$ columns in the 1D and 2D boxes are to be finalized by plugging $\bx=\bx_*$ to obtain the outputs in the second column. Interested readers are encouraged to consult the visual representation of the kernels in Fig. 2 of \cite{schulz2018tutorial}.

%%%%%%%%%%%%%%%%%%%%%%%%%%%%%%%
%       Section 6
%%%%%%%%%%%%%%%%%%%%%%%%%%%%%%%
\section{Kernels for Smooth Function Approximations and Numerical Results}
We present GP performance results on 1D and 2D grids for smoothly varying differentiable function approximations. For all smooth function test cases in this section, we use the SE kernel on the blocky diamond GP stencils of three GP radii, $r_{gp} = 1, 2,$ and $3.$ Later in \cref{sec:gp_das}, we introduce a new GP kernel called the DAS (Discontinuous ArcSin) kernel to approximate discontinuous functions, which will be compared with the SE kernel and an existing neural network kernel to demonstrate its exceptional oscillation-control property at discontinuities.

%%%%%%%%%%%%%%%%%%%%%%%%%%%%%%%
%       Section 6.1
%%%%%%%%%%%%%%%%%%%%%%%%%%%%%%%
\subsection{Constrained Kernels}
To conduct the tasks in \cref{tab:GP_applications}, we first present the constrained kernels associated with different linear operators. These kernels are derived by applying linear operators $\mathcal{L}_1$ to $K(\bx, \bx')$ and  $\mathcal{L}_2$ to $K(\bx_*, \bx')$ as described in the caption of \cref{tab:GP_applications}. Nothing needs to be done to the SE kernel for the baseline point-evaluation operator PE, which is in \cref{eq:se}, and we do not repeat it here. In all cases, the input vector $[\bg]_h=g(\bx_h)=\mathcal{L}_1(f(\bx_h))$ takes a linear operator $\mathcal{L}_1$ of the pointwise function values, $[\bff]_h = f(\bx_h)$, sampled from the cell centers of the blocky square. For example, in the volume average to point conversion, $\mathbf{g}$ would represent the volume-averaged data which is supplied as an input to the calculation. The general form of the GP prediction with linear constraints is introduced in \cref{eq:GP_general}, and is shown here again:
\beq\label{eq:GP_general2}
p_{*}=\bz^T_* \cdot \bg = \btt_*^T \bfC^{-1}\cdot \bg = 
\Bigg(\mathcal{L}_2 \Big( K(\bx_{*},\bx') \Big)\Bigg)^T \Bigg(\mathcal{L}_1 \Big( K (\bx,\bx') \Big)\Bigg)^{-1} \cdot \bfg.
\eeq

%%%%%%%%%%%%%%%%%%%%%%%%%%%%%%%
%       Section 6.1.1
%%%%%%%%%%%%%%%%%%%%%%%%%%%%%%%
\subsubsection{Kernels for pointwise function values to a pointwise interpolation in 1D}\label{sec:pt-to-pt_1D}
The basic SE kernel $K_{\text{SE}}$ in \cref{eq:se} is all we need to calculate the baseline pointwise-to-pointwise function interpolation that takes an input data $\bff$ to predict $f(\bx_*)$, where both $\mathcal{L}_1$ and $\mathcal{L}_2$ are the point-evaluation operator PE. In this case, the GP prediction at $\bx_*$ is simply the updated posterior mean in \cref{eq:PostMean} using the SE kernel matrix, $\bK = \bK_{\text{SE}}$, i.e.,
\begin{equation}\label{eq:pt-to-pt}
    f(\bx_*) = \bz_*^T \cdot \bff, \quad \text{where}\;\; \bz^T_* = \bk_*^T \bK^{-1}.
\end{equation}

As an example, we present a grid convergence study to approximate the function $f(x) = e^{-x}\sin(4\pi x) \cos(2\pi x)$ on $[0, 1]$ discretized at $x=x_i$ on five grid resolutions, $N_x = 16, 32, 64, 128, 256$ in \cref{tab:pt-to-pt_1d}. GP takes the cell-centered function values $f(x_i)$ as input on three GP stencils of size $r_{gp}=1,2,3$ to make function value predictions at $x=x_{i+1/2}$. In all experimental grid convergence tests hereafter, we display the standard experimental order of convergence (EOC) rates defined by \footnote{Note that this EOC is  different from the effective order of convergence we used for the Taylor series analysis in \cref{eq:eoc}.}
\beq\label{eq:eoc_grid}
\mbox{EOC}  = \frac{\ln(E_c/E_r)}{\ln(2)},
\eeq
where $E_c$ and $E_r$ are the errors in the corresponding norm
on the coarse and the next refined resolutions
(e.g., $E_c$ on $32$ and $E_r$ on $64$),
respectively, calculated against the exact solution on each grid.
We see that EOCs are third, fifth, and seventh, respectively on $r_{gp}=1,2,3$, successfully following the anticipated order, $(2r_{gp}+1)$, on each grid resolution.

%%%%%%%%%%%%%%%%%%%%%%%%%%%%%%%
%       Table 3
%%%%%%%%%%%%%%%%%%%%%%%%%%%%%%%
\begin{table}[ht!]
    %\footnotesize
    \scriptsize
    \centering
    \caption{Grid convergence results for the 1D point-to-point interpolation in \cref{sec:pt-to-pt_1D}. GP calculations are performed using the SE kernel with $r_{gp}=1, 2, 3$ to approximate the function values $f(x_{i+1/2})$ from the cell center input values $f(x_i)$ for $f(x) = e^{-x}\sin(4\pi x) \cos(2\pi x)$ on $[0, 1]$. 
    We set $\ell =0.05$.}
    %%%
    \begin{tabular}{@{}cccc|ccc|cccc@{}}
        \toprule
        \multirow{2}{*}{Grid Res.} & \multicolumn{3}{c}{$r_{gp} = 1$} & \multicolumn{3}{c}{$r_{gp} = 2$} & \multicolumn{3}{c}{$r_{gp} = 3$} \\ 
        \cmidrule(lr){2-4} \cmidrule(lr){5-7} \cmidrule(lr){8-10} 
        & $L_1$ error &  $L_2$ error & $L_\infty$ error & $L_1$ error &  $L_2$ error & $L_\infty$ error & $L_1$ error &  $L_2$ error & $L_\infty$ error \\
        \midrule
        \( 16 \)   & 1.5755e-02 & 1.9291e-02 & 4.1273e-02 & 7.9488e-03 & 9.5539e-03 & 1.8090e-02 & 4.1997e-03 & 5.1116e-03 & 1.0107e-02 \\
        \( 32 \)   & 5.6083e-03 & 6.7936e-03 & 1.6785e-02 & 1.0475e-03 & 1.3935e-03 & 3.7604e-03 & 4.3879e-04 & 5.1092e-04 & 1.1524e-03 \\
        \( 64 \)  & 7.9883e-04 & 9.6524e-04 & 2.6346e-03 & 4.8332e-05 & 6.1595e-05 & 1.7997e-04 & 4.4746e-06 & 6.3025e-06 & 1.8003e-05 \\
        \( 128 \)  & 1.0100e-04 & 1.2283e-04 & 3.4421e-04 & 1.5971e-06 & 2.0460e-06 & 5.9998e-06 & 3.8455e-08 & 5.3534e-08 & 1.5710e-07 \\
        \( 256 \)  & 1.2651e-05 & 1.5437e-05 & 4.3386e-05 & 5.0499e-08 & 6.4622e-08 & 1.8882e-07 & 3.0713e-10 & 4.2503e-10 & 1.2402e-09 \\
        \midrule
        & $L_1$ EOC &  $L_2$ EOC & $L_\infty$ EOC & $L_1$ EOC &  $L_2$ EOC & $L_\infty$ EOC & $L_1$ EOC &  $L_2$ EOC & $L_\infty$ EOC \\
        \midrule
        \( 16 \)   & -- & -- & -- & -- & -- & -- & -- & -- & -- \\
        \( 32 \)   & 1.490 & 1.506 & 1.298 & 2.924 & 2.777 & 2.266 & 3.259 & 3.323 & 3.133 \\
        \( 64 \)  & 2.812 & 2.815 & 2.672 & 4.438 & 4.500 & 4.385 & 6.616 & 6.341 & 6.000 \\
        \( 128 \)  & 2.983 & 2.974 & 2.936 & 4.919 & 4.912 & 4.907 & 6.862 & 6.879 & 6.840 \\
        \( 256 \)  & 2.997 & 2.992 & 2.988 & 4.983 & 4.985 & 4.990 & 6.968 & 6.977 & 6.985 \\
        \bottomrule
    \end{tabular}	\label{tab:pt-to-pt_1d}
\end{table}

%%%%%%%%%%%%%%%%%%%%%%%%%%%%%%%
%       Section 6.1.2
%%%%%%%%%%%%%%%%%%%%%%%%%%%%%%%
\subsubsection{Kernels for pointwise function values to a pointwise interpolation in 2D}\label{sec:pt-to-pt_2D}
Similarly, the 2D pointwise interpolation uses the same SE kernel except that the kernel function $K_{\text{SE}}$ now takes the two-dimensional form with $\bx=(x,y), \bx'=(x',y') \in \mathbb{R}^2$,
\begin{equation}
    K_{\text{SE}} \left( \bx, \bx' \right) = e^{- \frac{\left(x - x'\right)^{2}}{2 \ell^{2}}} e^{- \frac{\left(y - y'\right)^{2}}{2 \ell^{2}}}.
    \label{eq:se_2D}
\end{equation}
Once a set of function values on a 2D GP stencil is flattened into a 1D vector, $\bff$, in an order consistent with the correlation order in the kernel matrix $\bK_{\text{SE}}$, the 2D calculation proceeds with the same expression as in \cref{eq:pt-to-pt}.

As in the 1D case in the previous section, we also present a 2D grid convergence study in \cref{tab:pt-to-pt_2d}. We discretize $f(x,y) = e^{-2x}\sin(4 \pi y) + x e^{-y}\cos(2 \pi x)$  at cell centers $(x_i, y_j)$ over the square domain $[0, 1]\times[0,1]$. Three GP calculations on the blocky stencil with $r_{gp}=1,2,3$ task to predict the function values at the cell corner locations $(x_{i+1/2},y_{j+1/2})$ on five grid resolutions, $N_x = 16, 32, 64, 128, 256$. Again, the 2D EOC rates are third, fifth, and seventh, respectively on $r_{gp}=1,2,3$, satisfactorily demonstrating the expected rates of convergence in all cases. In particular, the experimental result with $r_{gp}=2$ is consistent with the anticipated and \textit{effective} target order study for the blocky diamond stencil in \cref{tab:stencil_convergences}. As mentioned, we set the blocky diamond GP stencil as our default 2D stencil in all 2D calculations hereafter.

%%%%%%%%%%%%%%%%%%%%%%%%%%%%%%%
%       Table 4
%%%%%%%%%%%%%%%%%%%%%%%%%%%%%%%
\begin{table}[ht!]
    %\footnotesize
    \scriptsize
        \centering
        \caption{
        Grid convergence results for the 2D point-to-point interpolation in \cref{sec:pt-to-pt_2D}. GP calculations are performed using the SE kernel with $r_{gp}=1, 2, 3$ to approximate the function values $f(x_{i+1/2},y_{j+1/2})$ from the cell center input values $f(x_i, y_j)$ for $f(x,y) = e^{-2x}\sin(4 \pi y) + x e^{-y}\cos(2 \pi x)$ on $[0, 1]\times[0,1]$. As discussed, we set the blocky diamond stencil as our default 2D stencil configuration. We choose $\ell =0.05$ for the length hyperparameter.
}
   	%%%
        \begin{tabular}{@{}cccc|ccc|cccc@{}}
            \toprule
            \multirow{2}{*}{Grid Res.} & \multicolumn{3}{c}{$r_{gp} = 1$} & \multicolumn{3}{c}{$r_{gp} = 2$} & \multicolumn{3}{c}{$r_{gp} = 3$} \\ 
            \cmidrule(lr){2-4} \cmidrule(lr){5-7} \cmidrule(lr){8-10} 
            & $L_1$ error &  $L_2$ error & $L_\infty$ error & $L_1$ error &  $L_2$ error & $L_\infty$ error & $L_1$ error &  $L_2$ error & $L_\infty$ error \\
            \midrule
            \( 16 \)   & 4.1324e-02 & 4.9148e-02 & 1.1271e-01 & 1.3467e-02 & 1.5832e-02 & 3.8873e-02 & 4.5533e-03 & 5.4040e-03 & 1.3892e-02 \\
            \( 32 \)   & 8.2994e-03 & 9.9775e-03 & 2.5316e-02 & 2.2561e-03 & 2.6967e-03 & 6.7348e-03 & 8.9778e-04 & 1.0605e-03 & 2.3860e-03 \\
            \( 64 \)  & 1.0237e-03 & 1.2386e-03 & 3.3654e-03 & 8.2938e-05 & 9.9982e-05 & 2.8072e-04 & 1.0241e-05 & 1.2241e-05 & 3.1432e-05 \\
            \( 128 \)  & 1.2581e-04 & 1.5315e-04 & 4.2373e-04 & 2.5631e-06 & 3.1014e-06 & 8.3960e-06 & 8.1480e-08 & 9.7758e-08 & 2.5652e-07 \\
            \( 256 \)  & 1.5589e-05 & 1.9035e-05 & 5.2948e-05 & 7.9461e-08 & 9.6359e-08 & 2.6240e-07 & 6.3098e-10 & 7.5861e-10 & 2.0054e-09 \\

            \midrule
            & $L_1$ EOC &  $L_2$ EOC & $L_\infty$ EOC & $L_1$ EOC &  $L_2$ EOC & $L_\infty$ EOC & $L_1$ EOC &  $L_2$ EOC & $L_\infty$ EOC \\
            \midrule
            \( 16 \)   & -- & -- & -- & -- & -- & -- & -- & -- & -- \\
            \( 32 \)   & 2.32 & 2.30 & 2.15 & 2.58 & 2.55 & 2.53 & 2.34 & 2.35 & 2.54 \\
            \( 64 \)  & 3.02 & 3.01 & 2.91 & 4.77 & 4.75 & 4.58 & 6.45 & 6.44 & 6.25 \\
            \( 128 \)  & 3.02 & 3.02 & 2.99 & 5.02 & 5.01 & 5.06 & 6.97 & 6.97 & 6.94 \\
            \( 256 \)  & 3.01 & 3.01 & 3.00 & 5.01 & 5.01 & 5.00 & 7.01 & 7.01 & 7.00 \\
            \bottomrule
        \end{tabular} \label{tab:pt-to-pt_2d}
    \end{table}

%%%%%%%%%%%%%%%%%%%%%%%%%%%%%%%
%       Section 6.1.3
%%%%%%%%%%%%%%%%%%%%%%%%%%%%%%%
\subsubsection{Kernels for pointwise function values to a derivative in 1D} \label{sec:pt-to-der_1D}
The GP prediction from pointwise input values to a derivative takes $\mathcal{L}_1$=PE and $\mathcal{L}_2$ is to be a relevant derivative operator at $x=x_*$. In 1D, we get
\begin{equation}
    \mathcal{L}_2({K}_{\text{SE}}(x_*,x')) =\Bigg(\frac{d^m}{dx^m}{\mathbf{K}_{\text{SE}}(x,x')}\Bigg) \Bigg|_{x=x_*}.
\end{equation}
For brevity, we only show kernels for the first, second, and third derivatives here.
For the first derivative ($m=1$), we get
\begin{equation}\label{eq:pt-to-1stDer_1d}
\mathcal{L}_2({K}_{\text{SE}}(x_*,x')) = - \frac{\left(- 2 x_* + 2 x^\prime \right) e^{- \frac{\left(x_* - x^\prime\right)^{2}}{2 \ell^{2}}}}{2 \ell^{2}}.
\end{equation}
The second derivative kernel ($m=2$) takes of the form
\begin{equation}\label{eq:pt-to-2ndDer_1d}
\mathcal{L}_2({K}_{\text{SE}}(x_*,x')) = \frac{\left(-1 + \frac{\left(x_* - x^\prime\right)^{2}}{\ell^{2}}\right) e^{- \frac{\left(x_* - x^\prime\right)^{2}}{2 \ell^{2}}}}{\ell^{2}},
\end{equation}
and finally, for the third derivative ($m=3$), the kernel is given as
\begin{equation}\label{eq:pt-to-3rdDer_1d}
\mathcal{L}_2({K}_{\text{SE}}(x_*,x')) = \frac{\left(-3 + \frac{\left(x_* - x^\prime\right)^{2}}{\ell^{2}}\right) \left(x - x^\prime\right) e^{- \frac{\left(x_* - x^\prime\right)^{2}}{2 \ell^{2}}}}{\ell^{4}}.
\end{equation}

Numerical convergence results are shown in \cref{tab:pt-to-1stDer_1d,tab:pt-to-2ndDer-1d,tab:pt-to-3rdDer_1d}, respectively for the first, second, and third derivatives of the function $f(x) = e^{-x}\sin(4\pi x) \cos(2\pi x)$ defined on $[0, 1]$. As before, given the pointwise function values $f(x_i)$ at cell centers, we compute $f'(x_{i+1/2}), f''(x_{i+1/2})$, and $f'''(x_{i+1/2})$ at the cell interface locations using the corresponding kernel functions in \cref{eq:pt-to-1stDer_1d,eq:pt-to-2ndDer_1d,eq:pt-to-3rdDer_1d}.
As the grid refines, these test results satisfy the anticipated and effective convergence rates in \cref{tab:stencil_convergences} very well as the order of the derivative varies.

%%%%%%%%%%%%%%%%%%%%%%%%%%%%%%%
%       Table 5
%%%%%%%%%%%%%%%%%%%%%%%%%%%%%%%
\begin{table}[ht!]
    %\footnotesize
    \scriptsize    
    \centering
    \caption{Grid convergence results for the 1D point-to-1st derivative conversion in \cref{sec:pt-to-der_1D}. GP calculations are performed at cell interface locations $x_{i+1/2}$ using the kernel in \cref{eq:pt-to-1stDer_1d} with $r_{gp}=1, 2, 3$ and $\ell =0.05$ to approximate the function $f(x) = e^{-x}\sin(4\pi x) \cos(2\pi x)$ on $[0, 1]$. }
        \begin{tabular}{@{}cccc|ccc|cccc@{}}
        \toprule
        \multirow{2}{*}{Grid Res.} & \multicolumn{3}{c}{$r_{gp} = 1$} & \multicolumn{3}{c}{$r_{gp} = 2$} & \multicolumn{3}{c}{$r_{gp} = 3$} \\ 
        \cmidrule(lr){2-4} \cmidrule(lr){5-7} \cmidrule(lr){8-10} 
        & $L_1$ error &  $L_2$ error & $L_\infty$ error & $L_1$ error &  $L_2$ error & $L_\infty$ error & $L_1$ error &  $L_2$ error & $L_\infty$ error \\
        \midrule
        \( 16 \)   & 3.3539e-01 & 4.0471e-01 & 8.9270e-01 & 8.9040e-02 & 1.0535e-01 & 2.1358e-01 & 1.1327e-02 & 1.4364e-02 & 3.0059e-02 \\
        \( 32 \)   & 1.0599e-01 & 1.2858e-01 & 2.8091e-01 & 1.5199e-02 & 1.8424e-02 & 3.8738e-02 & 5.3990e-03 & 6.6940e-03 & 1.5706e-02 \\
        \( 64 \)  & 3.0356e-02 & 3.6988e-02 & 1.0172e-01 & 1.1218e-03 & 1.4032e-03 & 3.7392e-03 & 7.2870e-05 & 9.2416e-05 & 2.3222e-04 \\
        \( 128 \)  & 8.3399e-03 & 1.0109e-02 & 2.8394e-02 & 7.7643e-05 & 9.8382e-05 & 2.8879e-04 & 1.3063e-06 & 1.7915e-06 & 5.2338e-06 \\
        \( 256 \)  & 2.1284e-03 & 2.5927e-03 & 7.3070e-03 & 5.0474e-06 & 6.4345e-06 & 1.8915e-05 & 2.1498e-08 & 2.9792e-08 & 8.7917e-08 \\
        \midrule
        & $L_1$ EOC &  $L_2$ EOC & $L_\infty$ EOC & $L_1$ EOC &  $L_2$ EOC & $L_\infty$ EOC & $L_1$ EOC &  $L_2$ EOC & $L_\infty$ EOC \\
        \midrule
        \( 16 \)   & -- & -- & -- & -- & -- & -- & -- & -- & -- \\
        \( 32 \)   & 1.662 & 1.654 & 1.668 & 2.550 & 2.516 & 2.463 & 1.069 & 1.101 & 0.937 \\
        \( 64 \)  & 1.804 & 1.798 & 1.466 & 3.760 & 3.715 & 3.373 & 6.211 & 6.179 & 6.080 \\
        \( 128 \)  & 1.864 & 1.871 & 1.841 & 3.853 & 3.834 & 3.695 & 5.802 & 5.689 & 5.471 \\
        \( 256 \)  & 1.970 & 1.963 & 1.958 & 3.943 & 3.935 & 3.932 & 5.925 & 5.910 & 5.896 \\

        \bottomrule
    \end{tabular}\label{tab:pt-to-1stDer_1d}
\end{table}

%%%%%%%%%%%%%%%%%%%%%%%%%%%%%%%
%       Table 6
%%%%%%%%%%%%%%%%%%%%%%%%%%%%%%%
\begin{table}[ht!]
    %\footnotesize
    \scriptsize
        \centering
    \caption{Grid convergence results for the 1D point-to-2nd derivative conversion in \cref{sec:pt-to-der_1D} using the kernel in \cref{eq:pt-to-2ndDer_1d}. All other GP setups are the same as in \cref{tab:pt-to-1stDer_1d}.}
%%%%
    \begin{tabular}{@{}cccc|ccc|cccc@{}}
        \toprule
        \multirow{2}{*}{Grid Res.} & \multicolumn{3}{c}{$r_{gp} = 1$} & \multicolumn{3}{c}{$r_{gp} = 2$} & \multicolumn{3}{c}{$r_{gp} = 3$} \\ 
        \cmidrule(lr){2-4} \cmidrule(lr){5-7} \cmidrule(lr){8-10} 
        & $L_1$ error &  $L_2$ error & $L_\infty$ error & $L_1$ error &  $L_2$ error & $L_\infty$ error & $L_1$ error &  $L_2$ error & $L_\infty$ error \\
        \midrule
        \( 16 \)   & 2.9833e+01 & 3.6768e+01 & 7.8908e+01 & 1.6014e+01 & 1.9091e+01 & 3.6545e+01 & 8.7882e+00 & 1.0694e+01 & 2.0702e+01 \\
        \( 32 \)   & 4.5051e+01 & 5.4730e+01 & 1.3643e+02 & 9.3169e+00 & 1.2405e+01 & 3.3636e+01 & 4.0084e+00 & 4.6632e+00 & 1.0421e+01 \\
        \( 64 \)  & 2.5862e+01 & 3.1224e+01 & 8.5017e+01 & 1.7292e+00 & 2.2026e+00 & 6.4389e+00 & 1.6597e-01 & 2.3354e-01 & 6.6653e-01 \\
        \( 128 \)  & 1.3098e+01 & 1.5934e+01 & 4.4625e+01 & 2.2993e-01 & 2.9462e-01 & 8.6357e-01 & 5.7359e-03 & 7.9834e-03 & 2.3418e-02 \\
        \( 256 \)  & 6.5922e+00 & 8.0455e+00 & 2.2604e+01 & 2.9229e-02 & 3.7406e-02 & 1.0926e-01 & 1.8418e-04 & 2.5485e-04 & 7.4342e-04 \\
        \midrule
        & $L_1$ EOC &  $L_2$ EOC & $L_\infty$ EOC & $L_1$ EOC &  $L_2$ EOC & $L_\infty$ EOC & $L_1$ EOC &  $L_2$ EOC & $L_\infty$ EOC \\
        \midrule
        \( 16 \)   & -- & -- & -- & -- & -- & -- & -- & -- & -- \\
        \( 32 \)   & -0.595 & -0.574 & -0.790 & 0.781 & 0.622 & 0.120 & 1.133 & 1.197 & 0.990 \\
        \( 64 \)  & 0.801 & 0.810 & 0.682 & 2.430 & 2.494 & 2.385 & 4.594 & 4.320 & 3.967 \\
        \( 128 \)  & 0.982 & 0.971 & 0.930 & 2.911 & 2.902 & 2.898 & 4.855 & 4.871 & 4.831 \\
        \( 256 \)  & 0.990 & 0.986 & 0.981 & 2.976 & 2.977 & 2.982 & 4.961 & 4.969 & 4.977 \\

        \bottomrule
    \end{tabular}\label{tab:pt-to-2ndDer-1d}
\end{table}

%%%%%%%%%%%%%%%%%%%%%%%%%%%%%%%
%       Table 7
%%%%%%%%%%%%%%%%%%%%%%%%%%%%%%%
\begin{table}[ht!]
    %\footnotesize
    \scriptsize
\caption{Grid convergence results for the 1D point-to-3rd derivative conversion in \cref{sec:pt-to-der_1D} using the kernel in \cref{eq:pt-to-3rdDer_1d}. All other GP setups are the same as in \cref{tab:pt-to-1stDer_1d}.}
%%%
    \centering
    \begin{tabular}{@{}cccc|ccc|cccc@{}}
        \toprule
        \multirow{2}{*}{Grid Res.} & \multicolumn{3}{c}{$r_{gp} = 1$} & \multicolumn{3}{c}{$r_{gp} = 2$} & \multicolumn{3}{c}{$r_{gp} = 3$} \\ 
        \cmidrule(lr){2-4} \cmidrule(lr){5-7} \cmidrule(lr){8-10} 
        & $L_1$ error &  $L_2$ error & $L_\infty$ error & $L_1$ error &  $L_2$ error & $L_\infty$ error & $L_1$ error &  $L_2$ error & $L_\infty$ error \\
        \midrule
        \( 16 \)   & 2.0660e+03 & 2.4861e+03 & 5.5018e+03 & 5.6927e+02 & 6.7278e+02 & 1.3753e+03 & 7.0796e+01 & 8.7636e+01 & 1.8277e+02 \\
        \( 32 \)   & 2.5204e+03 & 3.0602e+03 & 6.7207e+03 & 4.0375e+02 & 4.8935e+02 & 1.0304e+03 & 1.4836e+02 & 1.8381e+02 & 4.3101e+02 \\
        \( 64 \)  & 2.9141e+03 & 3.5502e+03 & 9.7604e+03 & 1.1975e+02 & 1.4976e+02 & 3.9880e+02 & 8.0585e+00 & 1.0220e+01 & 2.5661e+01 \\
        \( 128 \)  & 3.2348e+03 & 3.9209e+03 & 1.1012e+04 & 3.3467e+01 & 4.2403e+01 & 1.2447e+02 & 5.8337e-01 & 7.9997e-01 & 2.3371e+00 \\
        \( 256 \)  & 3.3232e+03 & 4.0481e+03 & 1.1409e+04 & 8.7568e+00 & 1.1163e+01 & 3.2816e+01 & 3.8640e-02 & 5.3548e-02 & 1.5802e-01 \\
        \midrule
        & $L_1$ EOC &  $L_2$ EOC & $L_\infty$ EOC & $L_1$ EOC &  $L_2$ EOC & $L_\infty$ EOC & $L_1$ EOC &  $L_2$ EOC & $L_\infty$ EOC \\
        \midrule
        \( 16 \)   & -- & -- & -- & -- & -- & -- & -- & -- & -- \\
        \( 32 \)   & -0.287 & -0.300 & -0.289 & 0.496 & 0.459 & 0.417 & -1.067 & -1.069 & -1.238 \\
        \( 64 \)  & -0.209 & -0.214 & -0.538 & 1.753 & 1.708 & 1.369 & 4.202 & 4.169 & 4.070 \\
        \( 128 \)  & -0.151 & -0.143 & -0.174 & 1.839 & 1.820 & 1.680 & 3.788 & 3.675 & 3.457 \\
        \( 256 \)  & -0.039 & -0.046 & -0.051 & 1.934 & 1.925 & 1.923 & 3.916 & 3.901 & 3.887 \\

        \bottomrule
    \end{tabular}\label{tab:pt-to-3rdDer_1d}
\end{table}

%%%%%%%%%%%%%%%%%%%%%%%%%%%%%%%
%       Section 6.1.4
%%%%%%%%%%%%%%%%%%%%%%%%%%%%%%%
\subsubsection{Kernels for pointwise function values to a derivative in 2D} \label{sec:pt-to-der_2D}
In a similar fashion, kernels for derivatives in 2D can be obtained by considering $\mathcal{L}_1$=PE and $\mathcal{L}_2$ is a respective derivative operator at $\bx=(x,y)$, 
followed by the derivative evaluation at $\bx=\bx_*=(x_{*},y_{*})$. 
A general form is written as
\begin{equation}
    \mathcal{L}_2({K}_{\text{SE}}(\bx_*,\bx')) = \Bigg(\frac{\partial^{m+n}}{\partial y^n \partial x^m} {{K}_{\text{SE}}(\bx,\bx')} \Bigg) \Bigg|_{\bx=\bx_*}.
\end{equation}
We only show a couple of them here for exposition purposes. The kernel for the first-order partial derivative in $x$ is shown to be
\begin{equation}
  \frac{\partial}{\partial x} {{K}_{\text{SE}}(\bx_*,\bx')} =
  - \frac{\left(- 2 x_* + 2 x'\right) e^{- \frac{\left(x_* - x'\right)^{2}}{2 \ell^{2}}} e^{- \frac{\left(y_* - y'\right)^{2}}{2 \ell^{2}}}}{2 \ell^{2}}.
\label{eq:se_2d_ddx}
\end{equation}
Similarly, the kernel for $\partial_y$ becomes
%%%%%%%
\begin{equation}
  \frac{\partial}{\partial y} {K}_{\text{SE}}(\bx_*,\bx')= 
  - \frac{\left(- 2 y_* + 2 y'\right) e^{- \frac{\left(x_* - x'\right)^{2}}{2 \ell^{2}}} e^{- \frac{\left(y_* - y'\right)^{2}}{2 \ell^{2}}}}{2 \ell^{2}}.
\label{eq:se_2d_ddy}
\end{equation}
The kernel for the cross derivative $\partial_{xy}$ is
\begin{equation}
  \frac{\partial}{\partial x} \frac{\partial}{\partial y} {K}_{\text{SE}}(\bx_*,\bx')= 
  \frac{\left(- 2 x_* + 2 x'\right) \left(- 2 y_* + 2 y'\right) e^{- \frac{\left(x_* - x'\right)^{2}}{2 \ell^{2}}} e^{- \frac{\left(y_* - y'\right)^{2}}{2 \ell^{2}}}}{4 \ell^{4}},
\label{eq:se_2d_dxdy}
\end{equation}
and lastly the kernel for $\partial_{xx}$ is
%%%%%%% 
\begin{equation}
  \frac{\partial^2}{\partial x^2} {{K}_{\text{SE}}(\bx_*,\bx')} =
\frac{
\left(-1 + \frac{\left(x_* - x^\prime\right)^{2}}{\ell^{2}}\right) 
e^{- \frac{\left(x_* - x^\prime\right)^{2}}{2 \ell^{2}}} 
e^{- \frac{\left(y_* - y'\right)^{2}}{2 \ell^{2}}}
}
{\ell^{2}}.
\label{eq:se_2d_dxdx}
\end{equation}
Other higher-order derivatives can be readily obtained in a similar fashion.

\begin{figure}[H]
\centering
\includegraphics[width=0.95\textwidth]{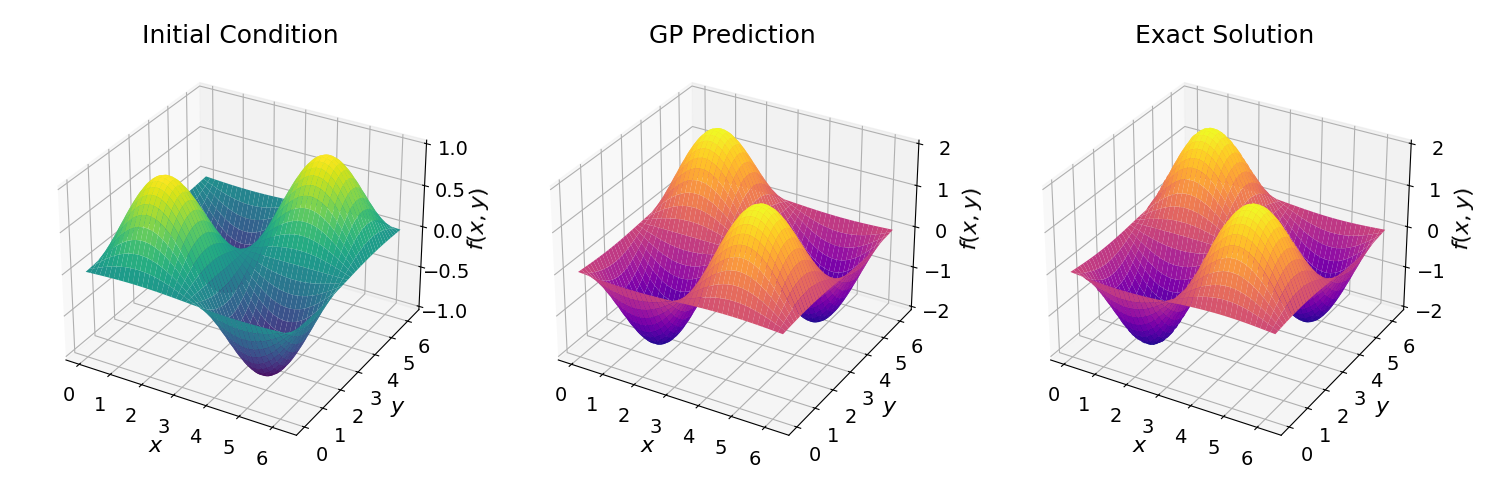}  
\caption{GP's prediction of $\nabla^2 \sin(x) \sin(y)$ on $[0,2\pi] \times [0, 2\pi]$ using a $128\times 128$ grid resolution with $r_{gp}=2$ and $\ell={\pi}/{10}$. The GP result shows a great visual agreement with the exact Laplacian calculation. The convergence study for this case is shown in \cref{table:laplacian_2d}.}
\label{fig:2d_lap}
\end{figure}

As a proof of concept, we show numerical results for computing a Laplacian operator of a function $\nabla^2 f=\frac{\partial^2 f}{\partial x^2}+\frac{\partial^2 f}{\partial y^2}$ to demonstrate the flexibility of the GP method. In this example, we choose a simple function $f(x,y)=\sin(x)\sin(y)$ on a square domain $[0,2\pi]  \times [0, 2\pi] $ although any other smooth functions would work well too. 
The function, the GP prediction of the Laplacian at cell corners $(x_{i+1/2},y_{j+1/2})$, and the exact Laplacian $\nabla^2 f(x_{i+1/2},y_{j+1/2})$ are plotted in \cref{fig:2d_lap}, where the GP solution and the exact solutions match well visually. To quantify GP's results, we conducted a grid convergence study for this test in \cref{table:laplacian_2d}. In all test cases, GP solutions successfully converge on each grid resolution following the anticipated and effective convergence rates described in \cref{tab:stencil_convergences}.

%%%%%%%%%%%%%%%%%%%%%%%%%%%%%%%
%       Fig. 3
%%%%%%%%%%%%%%%%%%%%%%%%%%%%%%%

%%%%%%%%%%%%%%%%%%%%%%%%%%%%%%%
%       Table 8
%%%%%%%%%%%%%%%%%%%%%%%%%%%%%%%
\begin{table}[ht!]
    %\footnotesize
    \scriptsize
        \centering
          \caption{Grid convergence results for the 2D point-to-Laplacian conversion in \cref{sec:pt-to-der_2D}. GP considers to approximate          
          the Laplacian of the function $f(x,y) = \sin(x) \sin(y)$ on $[0, 2 \pi] \times [0, 2\pi]$ at $(x_{i+1/2},y_{j+1/2})$ by taking the function values $f(x_i,y_j)$ at cell centers as input. The prediction utilizes the kernel in \cref{eq:se_2d_dxdx} with $\ell = {\pi}/{10}$.}
           \label{table:laplacian_2d}
        \begin{tabular}{@{}cccc|ccc|cccc@{}}
            \toprule
            \multirow{2}{*}{Grid Res.} & \multicolumn{3}{c}{$r_{gp} = 1$} & \multicolumn{3}{c}{$r_{gp} = 2$} & \multicolumn{3}{c}{$r_{gp} = 3$} \\ 
            \cmidrule(lr){2-4} \cmidrule(lr){5-7} \cmidrule(lr){8-10} 
            & $L_1$ error &  $L_2$ error & $L_\infty$ error & $L_1$ error &  $L_2$ error & $L_\infty$ error & $L_1$ error &  $L_2$ error & $L_\infty$ error \\
            \midrule
            \( 16 \)   & 1.1523e+02 & 2.1272e+01 & 5.9043e+00 & 5.5807e+01 & 1.0371e+01 & 3.0678e+00 & 1.7191e+01 & 3.0400e+00 & 6.8892e-01 \\
            \( 32 \)   & 8.3384e+01 & 1.4699e+01 & 3.6491e+00 & 3.0797e+01 & 5.4303e+00 & 1.3420e+00 & 1.3309e+01 & 2.3489e+00 & 5.7482e-01 \\
            \( 64 \)  & 3.8632e+01 & 6.8245e+00 & 1.6402e+00 & 3.8864e+00 & 6.8516e-01 & 1.6355e-01 & 5.3657e-01 & 9.4783e-02 & 2.2545e-02 \\
            \( 128 \)  & 1.8257e+01 & 3.2293e+00 & 7.7024e-01 & 4.5357e-01 & 8.0228e-02 & 1.9171e-02 & 1.6369e-02 & 2.8949e-03 & 6.7924e-04 \\
            \( 256 \)  & 9.0985e+00 & 1.6068e+00 & 3.7286e-01 & 5.6298e-02 & 9.9525e-03 & 2.3080e-03 & 5.0524e-04 & 8.9238e-05 & 2.0536e-05 \\

            \midrule
            & $L_1$ EOC &  $L_2$ EOC & $L_\infty$ EOC & $L_1$ EOC &  $L_2$ EOC & $L_\infty$ EOC & $L_1$ EOC &  $L_2$ EOC & $L_\infty$ EOC \\
            \midrule
            \( 16 \)   & -- & -- & -- & -- & -- & -- & -- & -- & -- \\
            \( 32 \)   & 0.47 & 0.53 & 0.69 & 0.86 & 0.93 & 1.19 & 0.37 & 0.37 & 0.26 \\
            \( 64 \)  & 1.11 & 1.11 & 1.15 & 2.99 & 2.99 & 3.04 & 4.63 & 4.63 & 4.67 \\
            \( 128 \)  & 1.08 & 1.08 & 1.09 & 3.10 & 3.09 & 3.09 & 5.03 & 5.03 & 5.05 \\
            \( 256 \)  & 1.00 & 1.01 & 1.05 & 3.01 & 3.01 & 3.05 & 5.02 & 5.02 & 5.05 \\

            \bottomrule
        \end{tabular}
    \end{table}

%%%%%%%%%%%%%%%%%%%%%%%%%%%%%%%
%       Section 6.1.5
%%%%%%%%%%%%%%%%%%%%%%%%%%%%%%%
\subsubsection{Kernels for pointwise function values to a volume average value in 1D} \label{sec:pt-to-va_1D}
As the input values are pointwise, we choose $\mathcal{L}_1$=PE to correlate pointwise values at $x$ and $x'$, while we set $\mathcal{L}_2$ to be a volume-averaging integral operator over $x$, followed by the integral evaluation at $x=x_*$. That is,
\begin{equation}
    \mathcal{L}_2({K}_{\text{SE}}(x_*,x')) =
    \Bigg (\frac{1}{\Delta x}\int_{I_{*}} 
    {K_{\text{SE}}(x,x')}dx \Bigg) \bigg|_{x=x_*}.
    \label{eq:va-1D}
\end{equation}
Here, the operation of $\mathcal{L}_2$ can be understood as a bridging step that converts the pointwise values to the target volume average value at $x=x_*$ by considering a Bayesian inference between the pointwise input data at each $x'$ and the sought volume-averaging evaluation at $x=x_*$. Unlike other differential or point-evaluation operations where the point $x_*$ could be chosen from the interior or boundaries of a cell $I_*$, we set $x_*$ to be a cell center location to make the volume-averaging meaningful in our calculations. The same is true for the 2D volume averages in the next subsection.
With this choice of $x_*$ and $I_*=[x_* - \Delta x / 2, x_* + \Delta x / 2]$, \cref{eq:va-1D} is written in terms of the error function by,
\begin{align}
\mathcal{L}_2({K}_{\text{SE}}(x_*,x')) = 
&\sqrt{\frac{\pi}{2}}\frac {\ell }{{\Delta x}} 
\left( 
\operatorname{erf} \Bigg( \frac{\big( x_* + {\Delta x}/{2} - x'\big)}{\sqrt{2} \ell} \Bigg) - 
\operatorname{erf} \Bigg( \frac{\big( x_* - {\Delta x}/{2} - x'\big)}{\sqrt{2} \ell} \Bigg) 
\right) \nonumber \\
=&\sqrt{\frac{\pi}{2}}\frac {\ell }{{\Delta x}} 
\left( 
\operatorname{erf} \Bigg( \frac{\big( \Delta_{x^* x'} + 1/2\big)}{\sqrt{2} \ell / \Delta x} \Bigg) - 
\operatorname{erf} \Bigg( \frac{\big( \Delta_{x^* x'} - 1/2\big)}{\sqrt{2} \ell / \Delta x} \Bigg) 
\right),
\label{eq:va_1D_err}
\end{align}
where $\Delta_{x^* x'} = {(x_{*}-x')}/{\Delta x}$.

Since the output is the integral average, $\frac{1}{\Delta x}\int_{I_{*}} f(x) dx$, the grid $\Delta x$ from the integration cancels with the averaging $\Delta x$. Hence, for each $r_{gp}$, the anticipated convergence rate is the same as the $(2r_{gp}+1)$ rate in the baseline pointwise to pointwise conversion in \cref{sec:pt-to-pt_1D} and \cref{tab:stencil_convergences}. Experimental GP results are collected in \cref{tab:Pt_to_VolAve_1D}, demonstrating the anticipated third, fifth, and seventh-order convergence rates on $r_{gp}=1,2,3$, respectively. 

%%%%%%%%%%%%%%%%%%%%%%%%%%%%%%%
%       Table 8
%%%%%%%%%%%%%%%%%%%%%%%%%%%%%%%
\begin{table}[ht!]
    %\footnotesize
    \scriptsize
    \centering
    \caption{Grid convergence results for the 1D point-to-volume average conversion in \cref{sec:pt-to-va_1D}. 
    Input values are $f(x_i)$ for the function $f(x) = e^{-x}\sin(4\pi x) \cos(2\pi x)$ on $[0, 1]$ and GP predicts $\frac{1}{\Delta x}\int_{I_i}f(x)dx$ at $x_i$ using the kernel in \cref{eq:va_1D_err} with $\ell =0.05$.}
        \begin{tabular}{@{}cccc|ccc|cccc@{}}
        \toprule
        \multirow{2}{*}{Grid Res.} & \multicolumn{3}{c}{$r_{gp} = 1$} & \multicolumn{3}{c}{$r_{gp} = 2$} & \multicolumn{3}{c}{$r_{gp} = 3$} \\ 
        \cmidrule(lr){2-4} \cmidrule(lr){5-7} \cmidrule(lr){8-10} 
        & $L_1$ error &  $L_2$ error & $L_\infty$ error & $L_1$ error &  $L_2$ error & $L_\infty$ error & $L_1$ error &  $L_2$ error & $L_\infty$ error \\
        \midrule
        \( 16 \)   & 1.0392e-02 & 1.2708e-02 & 2.7168e-02 & 5.1714e-03 & 6.2267e-03 & 1.1763e-02 & 2.7103e-03 & 3.2999e-03 & 6.5516e-03 \\
        \( 32 \)   & 3.7056e-03 & 4.4867e-03 & 1.1063e-02 & 6.7677e-04 & 9.0023e-04 & 2.4265e-03 & 2.8178e-04 & 3.2817e-04 & 7.4173e-04 \\
        \( 64 \)  & 5.3047e-04 & 6.4110e-04 & 1.7507e-03 & 3.1409e-05 & 4.0033e-05 & 1.1696e-04 & 2.8846e-06 & 4.0638e-06 & 1.1610e-05 \\
        \( 128 \)  & 6.7270e-05 & 8.1796e-05 & 2.2926e-04 & 1.0401e-06 & 1.3324e-06 & 3.9077e-06 & 2.4848e-08 & 3.4594e-08 & 1.0153e-07 \\
        \( 256 \)  & 8.4310e-06 & 1.0287e-05 & 2.8916e-05 & 3.2908e-08 & 4.2110e-08 & 1.2305e-07 & 1.9858e-10 & 2.7481e-10 & 8.0197e-10 \\
        \midrule
        & $L_1$ EOC &  $L_2$ EOC & $L_\infty$ EOC & $L_1$ EOC &  $L_2$ EOC & $L_\infty$ EOC & $L_1$ EOC &  $L_2$ EOC & $L_\infty$ EOC \\
        \midrule
        \( 16 \)   & -- & -- & -- & -- & -- & -- & -- & -- & -- \\
        \( 32 \)   & 1.488 & 1.502 & 1.296 & 2.934 & 2.790 & 2.277 & 3.266 & 3.330 & 3.143 \\
        \( 64 \)  & 2.804 & 2.807 & 2.660 & 4.429 & 4.491 & 4.375 & 6.610 & 6.335 & 5.997 \\
        \( 128 \)  & 2.979 & 2.970 & 2.933 & 4.916 & 4.909 & 4.904 & 6.859 & 6.876 & 6.837 \\
        \( 256 \)  & 2.996 & 2.991 & 2.987 & 4.982 & 4.984 & 4.989 & 6.967 & 6.976 & 6.984 \\

        \bottomrule
    \end{tabular} \label{tab:Pt_to_VolAve_1D}
\end{table}

%%%%%%%%%%%%%%%%%%%%%%%%%%%%%%%
%       Section 6.1.6
%%%%%%%%%%%%%%%%%%%%%%%%%%%%%%%
\subsubsection{Kernels for pointwise function values to a volume average value in 2D} \label{sec:pt-to-va_2D}
Similar to the 1D case, GP converts pointwise function input values to predict a volume average value at $\bx=\bx_*=(x_*,y_*)$ in 2D using $\mathcal{L}_1$=PE and 
\begin{equation}
    \mathcal{L}_2({K}_{\text{SE}}(\bx_*,\bx')) =
    \Bigg (\frac{1}{\Delta \bx}\int_{I_{*}\times J_{*}} {K_{\text{SE}}(\bx,\bx')}d\bx \Bigg) \bigg|_{\bx=\bx_*}=
    \Bigg (\frac{1}{\Delta x \Delta y}\int_{I_{*}} \int_{J_{*}} {K_{\text{SE}}(\bx,\bx')}dx dy \Bigg) \bigg|_{\bx=\bx_*}.
    \label{eq:va_2d}
\end{equation}
Again, the $\mathcal{L}_2$ operation channels the pointwise input values to the 2D volume average value at $\bx_*$ by constructing a correlation between the two different data type values, i.e., the pointwise inputs sampled from cell center points $\bx'$ and the sought volume average value anchored at the point $\bx_*$ centered in a 2D rectangular cell $I_* \times J_* = [x_* - \Delta x / 2, x_* + \Delta x / 2] \times [y_* - \Delta y / 2, y_* + \Delta y / 2]$.
Due to the dimensional factorization property in the SE kernel in \cref{eq:se_arbdim}, the 2D version of the volume-averaging kernel in \cref{eq:va_2d} can be easily written as
\begin{equation}
\mathcal{L}_2({K}_{\text{SE}}(\bx_*,\bx')) = 
\prod_{d=x,y} 
\sqrt{\frac{\pi}{2}}\frac {\ell }{{\Delta d}} 
\left( 
\operatorname{erf} \Bigg( \frac{\big( \Delta_{\bx^* \bx'}^d + 1/2\big)}{\sqrt{2} \ell / \Delta d} \Bigg) - 
\operatorname{erf} \Bigg( \frac{\big( \Delta_{\bx^* \bx'}^d - 1/2\big)}{\sqrt{2} \ell / \Delta d} \Bigg) 
\right),
\label{eq:va_2D_err}
\end{equation}
where $\Delta_{\bx^* \bx'}^d = {\be_d \cdot (\bx_{*}-\bx')}/{\Delta d}$ with $\be_d$ being the unit vector in each $d$-direction. 
Note that \cref{eq:va_1D_err,eq:va_2D_err} recover the $\btt_*$ kernel form in \cite{bourgeois2022gp}.
As in the 1D case, GP converges to the anticipated third, fifth, and seventh-order solutions in 2D as well, as shown in \cref{tab:pt_to_voleave_2D}.

%%%%%%%%%%%%%%%%%%%%%%%%%%%%%%%
%       Table 10
%%%%%%%%%%%%%%%%%%%%%%%%%%%%%%%
\begin{table}[ht!]
    %\footnotesize
    \scriptsize
        \centering
    \caption{Grid convergence results for the 2D point-to-volume average conversion in \cref{sec:pt-to-va_2D}. 
    Input values are $f(x_i,y_j)$ for the function $f(x,y) = e^{-2x}\sin(4 \pi y) + x e^{-y}\cos(2 \pi x)$ on $[0, 1] \times [0,1]$ and GP predicts $\frac{1}{\Delta x \Delta y}\int_{I_i}\int_{I_j}f(x,y)dxdy$ at $(x_i,y_j)$ using the kernel in \cref{eq:va_2D_err} with $\ell =0.05$.}
        \begin{tabular}{@{}cccc|ccc|cccc@{}}
            \toprule
            \multirow{2}{*}{Grid Res.} & \multicolumn{3}{c}{$r_{gp} = 1$} & \multicolumn{3}{c}{$r_{gp} = 2$} & \multicolumn{3}{c}{$r_{gp} = 3$} \\ 
            \cmidrule(lr){2-4} \cmidrule(lr){5-7} \cmidrule(lr){8-10} 
            & $L_1$ error &  $L_2$ error & $L_\infty$ error & $L_1$ error &  $L_2$ error & $L_\infty$ error & $L_1$ error &  $L_2$ error & $L_\infty$ error \\
            \midrule
            \( 16 \)   & 2.7145e-02 & 3.2285e-02 & 7.3896e-02 & 8.5402e-03 & 1.0034e-02 & 2.4582e-02 & 2.9554e-03 & 3.5202e-03 & 9.2304e-03 \\
            \( 32 \)   & 5.5162e-03 & 6.6279e-03 & 1.6774e-02 & 1.4634e-03 & 1.7485e-03 & 4.3727e-03 & 5.7681e-04 & 6.8143e-04 & 1.5310e-03 \\
            \( 64 \)  & 6.8251e-04 & 8.2504e-04 & 2.2399e-03 & 5.4005e-05 & 6.5078e-05 & 1.8207e-04 & 6.6111e-06 & 7.9002e-06 & 2.0278e-05 \\
            \( 128 \)  & 8.3890e-05 & 1.0208e-04 & 2.8236e-04 & 1.6715e-06 & 2.0222e-06 & 5.4755e-06 & 5.2676e-08 & 6.3195e-08 & 1.6582e-07 \\
            \( 256 \)  & 1.0394e-05 & 1.2689e-05 & 3.5295e-05 & 5.1815e-08 & 6.2829e-08 & 1.7111e-07 & 4.0878e-10 & 4.9126e-10 & 1.2968e-09 \\

            \midrule
            & $L_1$ EOC &  $L_2$ EOC & $L_\infty$ EOC & $L_1$ EOC &  $L_2$ EOC & $L_\infty$ EOC & $L_1$ EOC &  $L_2$ EOC & $L_\infty$ EOC \\
            \midrule
            \( 16 \)   & -- & -- & -- & -- & -- & -- & -- & -- & -- \\
            \( 32 \)   & 2.30 & 2.28 & 2.14 & 2.54 & 2.52 & 2.49 & 2.36 & 2.37 & 2.59 \\
            \( 64 \)  & 3.01 & 3.01 & 2.90 & 4.76 & 4.75 & 4.59 & 6.45 & 6.43 & 6.24 \\
            \( 128 \)  & 3.02 & 3.01 & 2.99 & 5.01 & 5.01 & 5.06 & 6.97 & 6.97 & 6.93 \\
            \( 256 \)  & 3.01 & 3.01 & 3.00 & 5.01 & 5.01 & 5.00 & 7.01 & 7.01 & 7.00 \\
            \bottomrule
        \end{tabular}\label{tab:pt_to_voleave_2D}
    \end{table}

%%%%%%%%%%%%%%%%%%%%%%%%%%%%%%%
%       Section 6.1.7
%%%%%%%%%%%%%%%%%%%%%%%%%%%%%%%
\subsubsection{Kernels for volume average values to a pointwise value in 1D} \label{sec:va-to-pt_1D}
The conversion from volume averages at cell centers to a pointwise value at $x_*$ is one of the crucial building blocks in finite volume methods. 
The most common example is found in high-order reconstruction schemes that aim to achieve high-order accurate values called the Riemann states at cell interfaces, $x_*=x_{i\pm 1/2}$. Using GP, such a conversion can be formulated by the two kernels in \cref{eq:kernel_C,eq:kernel_t} for an input vector $\bg$, whose $h$th component is a 1D volume-averaged quantity $\frac{1}{\Delta x} \int_{I_h} f(x) dx$ of a function $f$. These volume average values between $x$ and $x'$ are correlated by $\mathcal{L}_1(K_{\text{SE}}(x,x'))$ that takes the volume-averaging integration of $K_{\text{SE}}(x,x')$ at both $x$ and $x'$,
%%%%%%%
\begin{align}
  \label{eq:va_SE_1D}
  \mathcal{L}_1(K_{\text{SE}}(x,x')) = 
  \sqrt{\pi}\left ( \ \frac{\ell}{\Delta x} \right)^2 & 
  \left \{ \left (
     \frac{\Delta_{xx'}+1}{\sqrt{2}\ell/\Delta x}\erf{\frac{\Delta_{xx'}+1}{\sqrt{2}\ell/\Delta x}}
   + \frac{\Delta_{xx'}-1}{\sqrt{2}\ell/\Delta x}\erf{\frac{\Delta_{xx'}-1}{\sqrt{2}\ell/\Delta x}}
   \right ) \right .
   \nonumber \\
  +  \frac{1}{\sqrt{\pi}} & \left . 
      \left (
              \expo{-\frac{(\Delta_{xx'}+1)^2}{2(\ell/\Delta x)^2}} + 
              \expo{-\frac{(\Delta_{xx'}-1)^2}{2(\ell/\Delta x)^2}} \right
      ) \right . \nonumber \\ 
      -2 & \left . 
      \left (
               \frac{\Delta_{xx'}}{\sqrt{2}\ell/\Delta x}\erf{\frac{\Delta_{xx'}}{\sqrt{2}\ell/\Delta x}}
            + \frac{1}{\sqrt{\pi}}\expo{-\frac{\Delta_{xx'}^2}{2(\ell/\Delta x)^2}}\right ) 
       \right \}.
\end{align}
These correlated volume averages are then channeled through the second operation by $\mathcal{L}_2(K_{\text{SE}}(x_*,x'))$ where the volume-averaging operation is kept at $x'$ while the pointwise evaluation is to take place at $x=x_*$. This is be done by finalizing the conversion process using \cref{eq:va_1D_err}.

Rigorously speaking, the $\mathcal{L}_2$ kernels are different in converting pointwise inputs to a volume average in \cref{sec:pt-to-va_1D} and volume average inputs to a pointwise output in this section. This clarification is noted in \cref{tab:GP_applications} making clear distinctions between $\frac{1}{\Delta \bx} \int  (\cdot) \; d\bx$ and $\frac{1}{\Delta \bx'} \int  (\cdot) \; d\bx'$ to indicate over which variable ($\bx$ or $\bx'$) the integration is computed. It is worth mentioning that there is no actual difference in the final expression of the kernel, though, since $\bx$ and $\bx'$ are interchangeable, and the SE kernel is symmetric positive definite. Lastly, the pointwise prediction at $x_*$ is obtained by combining the kernels in \cref{eq:va_SE_1D,eq:va_1D_err} according to \cref{eq:GP_general2}.

Similar to the case in \cref{sec:pt-to-va_1D}, we expect the volume average-to-pointwise conversion converges at $(2r_{gp}+1)$. Our experimental results confirm this convergence rate for each $r_{gp}=1,2,3$ in \cref{tab:va-to-pt_1d}.

%%%%%%%%%%%%%%%%%%%%%%%%%%%%%%%
%       Table 11
%%%%%%%%%%%%%%%%%%%%%%%%%%%%%%%
\begin{table}[ht!]
    %\footnotesize
    \scriptsize
    \centering
        \caption{Grid convergence results for the 1D volume-average-to-point conversion in \cref{sec:va-to-pt_1D}. Input values are the integral average values $\frac{1}{\Delta x}\int_{I_i} f(x) dx$ at cell centers $x_i$ and output values are pointwise function values $f(x_{i+1/2})$ at cell interface locations for $f(x) = e^{-x}\sin(4\pi x) \cos(2\pi x)$ on $[0, 1]$. GP utilizes the kernel in \cref{eq:va_SE_1D} with $\ell =0.05$.}
        \label{table:VolAve_to_Point_1D}
    \begin{tabular}{@{}cccc|ccc|cccc@{}}
        \toprule
        \multirow{2}{*}{Grid Res.} & \multicolumn{3}{c}{$r_{gp} = 1$} & \multicolumn{3}{c}{$r_{gp} = 2$} & \multicolumn{3}{c}{$r_{gp} = 3$} \\ 
        \cmidrule(lr){2-4} \cmidrule(lr){5-7} \cmidrule(lr){8-10} 
        & $L_1$ error &  $L_2$ error & $L_\infty$ error & $L_1$ error &  $L_2$ error & $L_\infty$ error & $L_1$ error &  $L_2$ error & $L_\infty$ error \\
        \midrule
        \( 16 \)   & 1.7407e-02 & 2.2542e-02 & 4.8096e-02 & 1.2685e-02 & 1.5502e-02 & 3.0334e-02 & 7.2806e-03 & 8.9043e-03 & 1.8025e-02 \\
        \( 32 \)   & 6.9198e-03 & 8.4268e-03 & 2.1101e-02 & 1.2698e-03 & 1.7192e-03 & 4.6781e-03 & 5.7232e-04 & 6.5120e-04 & 1.4616e-03 \\
        \( 64 \)  & 1.0496e-03 & 1.2688e-03 & 3.4631e-03 & 6.6081e-05 & 8.4426e-05 & 2.4717e-04 & 6.2253e-06 & 8.7481e-06 & 2.4844e-05 \\
        \( 128 \)  & 1.3419e-04 & 1.6327e-04 & 4.5752e-04 & 2.2506e-06 & 2.8857e-06 & 8.4621e-06 & 5.5469e-08 & 7.7300e-08 & 2.2659e-07 \\
        \( 256 \)  & 1.6858e-05 & 2.0574e-05 & 5.7815e-05 & 7.1677e-08 & 9.1745e-08 & 2.6803e-07 & 4.4789e-10 & 6.1992e-10 & 1.8082e-09 \\
        \midrule
        & $L_1$ EOC &  $L_2$ EOC & $L_\infty$ EOC & $L_1$ EOC &  $L_2$ EOC & $L_\infty$ EOC & $L_1$ EOC &  $L_2$ EOC & $L_\infty$ EOC \\
        \midrule
        \( 16 \)   & -- & -- & -- & -- & -- & -- & -- & -- & -- \\
        \( 32 \)   & 1.331 & 1.420 & 1.189 & 3.321 & 3.173 & 2.697 & 3.669 & 3.773 & 3.624 \\
        \( 64 \)  & 2.721 & 2.731 & 2.607 & 4.264 & 4.348 & 4.242 & 6.523 & 6.218 & 5.878 \\
        \( 128 \)  & 2.967 & 2.958 & 2.920 & 4.876 & 4.871 & 4.868 & 6.810 & 6.822 & 6.777 \\
        \( 256 \)  & 2.993 & 2.988 & 2.984 & 4.973 & 4.975 & 4.981 & 6.952 & 6.962 & 6.969 \\
        \bottomrule
    \end{tabular}\label{tab:va-to-pt_1d}
\end{table}

%%%%%%%%%%%%%%%%%%%%%%%%%%%%%%%
%       Section 6.1.8
%%%%%%%%%%%%%%%%%%%%%%%%%%%%%%%
\subsubsection{Kernels for volume average values to a pointwise value in 2D} \label{sec:va-to-pt_2D}
%$\frac{1}{\Delta\mathcal{V}}\int \bff \mathcal{V}$ to $f(\bx_*)$
The 1D calculation in the previous section is easily extended to 2D by combining $\mathcal{L}_2({K}_{\text{SE}}(\bx_*,\bx'))$ in \cref{eq:va_2D_err} and $\mathcal{L}_1({K}_{\text{SE}}(\bx,\bx'))$ defined by
\begin{align}
  \label{eq:SE-cov}
  \mathcal{L}_1(K_{\text{SE}}(\bx,\bx')) = 
  \prod_{d=x,y} \sqrt{\pi}\left ( \ \frac{\ell}{\Delta d} \right)^2 & 
  \left \{ \left (
     \frac{\Delta_{\bx\bx'}^d+1}{\sqrt{2}\ell/\Delta_d}\erf{\frac{\Delta_{\bx\bx'}^d+1}{\sqrt{2}\ell/\Delta d}}
   + \frac{\Delta_{\bx\bx'}^d-1}{\sqrt{2}\ell/\Delta_d}\erf{\frac{\Delta_{\bx\bx'}^d-1}{\sqrt{2}\ell/\Delta d}}
   \right ) \right .
   \nonumber \\
  +  \frac{1}{\sqrt{\pi}} & \left . 
      \left (
              \expo{-\frac{(\Delta_{\bx\bx'}^d+1)^2}{2(\ell/\Delta d)^2}} + 
              \expo{-\frac{(\Delta_{\bx\bx'}^d-1)^2}{2(\ell/\Delta d)^2}} \right
      ) \right . \nonumber \\ 
      -2 & \left . 
      \left (
               \frac{\Delta_{\bx\bx'}^d}{\sqrt{2}\ell/\Delta d}\erf{\frac{\Delta_{\bx\bx'}^d}{\sqrt{2}\ell/\Delta d}}
            + \frac{1}{\sqrt{\pi}}\expo{-\frac{(\Delta_{\bx\bx'}^d)^2}{2(\ell/\Delta d)^2}}\right ) 
       \right \}.
\end{align}
The resulting pointwise GP prediction has been used for designing two-dimensional high-order accurate finite volume Riemann states at $\bx_*=(x_{i\pm 1/2}, y_j)$ or $(x_{i}, y_{j\pm 1/2})$ in \cite{bourgeois2022gp}, directly converting the volume average quantities $\bg$ to a point value $p_*$ via the one-step dot product between the (pre-computed and saved) prediction vector $\bz_*^T$ and the input vector $\bg$ in \cref{eq:GP_general2}. This approach is an effective kernel-based alternative to the existing polynomial-based high-order multi-dimensional reconstruction schemes that incorporate high-order correction terms with Taylor series expansions \cite{mccorquodale2011high,buchmuller2014improved}.

In \cref{tab:va-to-pt_2d}, we once again confirm that the experimental convergence follows the $(2r_{gp}+1)$ rate for each GP radius, $r_{gp}=1,2,3$, similar to the results in \cref{sec:pt-to-va_2D}.

%%%%%%%%%%%%%%%%%%%%%%%%%%%%%%%
%       Table 12
%%%%%%%%%%%%%%%%%%%%%%%%%%%%%%%
\begin{table}[ht!]
    %\footnotesize
    \scriptsize
        \centering
      \caption{Grid convergence results for the 2D volume average-to-point conversion in \cref{sec:va-to-pt_2D}. Input values are the integral average values $\frac{1}{\Delta x \Delta y}\int_{I_i} \int_{I_j}f(x,y) dxdy$ at cell centers $(x_i,y_j)$ and output values are pointwise function values $f(x_{i+1/2},y_{i+1/2})$ at cell corner locations for $f(x,y) = e^{-2x}\sin(4 \pi y) + x e^{-y}\cos(2 \pi x) $ on $[0, 1]\times[0,1]$. GP utilizes the kernel in \cref{eq:SE-cov} with $\ell =0.05$.}
        \label{table:vole_ave_to_pt_2d}
        \begin{tabular}{@{}cccc|ccc|cccc@{}}
            \toprule
            \multirow{2}{*}{Grid Res.} & \multicolumn{3}{c}{$r_{gp} = 1$} & \multicolumn{3}{c}{$r_{gp} = 2$} & \multicolumn{3}{c}{$r_{gp} = 3$} \\ 
            \cmidrule(lr){2-4} \cmidrule(lr){5-7} \cmidrule(lr){8-10} 
            & $L_1$ error &  $L_2$ error & $L_\infty$ error & $L_1$ error &  $L_2$ error & $L_\infty$ error & $L_1$ error &  $L_2$ error & $L_\infty$ error \\
            \midrule
            \( 16 \)   & 4.7181e-02 & 5.6016e-02 & 1.3170e-01 & 1.7241e-02 & 2.0195e-02 & 5.0511e-02 & 7.9703e-03 & 9.4562e-03 & 2.3277e-02 \\
            \( 32 \)   & 1.0275e-02 & 1.2368e-02 & 3.1682e-02 & 2.8104e-03 & 3.3617e-03 & 8.3630e-03 & 1.0964e-03 & 1.2940e-03 & 2.8774e-03 \\
            \( 64 \)  & 1.3376e-03 & 1.6228e-03 & 4.4180e-03 & 1.1389e-04 & 1.3746e-04 & 3.8678e-04 & 1.4246e-05 & 1.7033e-05 & 4.3726e-05 \\
            \( 128 \)  & 1.6685e-04 & 2.0334e-04 & 5.6283e-04 & 3.6118e-06 & 4.3724e-06 & 1.1834e-05 & 1.1770e-07 & 1.4124e-07 & 3.7052e-07 \\
            \( 256 \)  & 2.0753e-05 & 2.5355e-05 & 7.0532e-05 & 1.1272e-07 & 1.3672e-07 & 3.7228e-07 & 9.1941e-10 & 1.1057e-09 & 2.9244e-09 \\

            \midrule
            & $L_1$ EOC &  $L_2$ EOC & $L_\infty$ EOC & $L_1$ EOC &  $L_2$ EOC & $L_\infty$ EOC & $L_1$ EOC &  $L_2$ EOC & $L_\infty$ EOC \\
            \midrule
            \( 16 \)   & -- & -- & -- & -- & -- & -- & -- & -- & -- \\
            \( 32 \)   & 2.20 & 2.18 & 2.06 & 2.62 & 2.59 & 2.59 & 2.86 & 2.87 & 3.02 \\
            \( 64 \)  & 2.94 & 2.93 & 2.84 & 4.63 & 4.61 & 4.43 & 6.27 & 6.25 & 6.04 \\
            \( 128 \)  & 3.00 & 3.00 & 2.97 & 4.98 & 4.97 & 5.03 & 6.92 & 6.91 & 6.88 \\
            \( 256 \)  & 3.01 & 3.00 & 3.00 & 5.00 & 5.00 & 4.99 & 7.00 & 7.00 & 6.99 \\

            \bottomrule
        \end{tabular}\label{tab:va-to-pt_2d}
        \end{table}

%%%%%%%%%%%%%%%%%%%%%%%%%%%%%%%
%       Section 6.1.9
%%%%%%%%%%%%%%%%%%%%%%%%%%%%%%%
\subsubsection{Kernels for volume-average values to a derivative in 2D} \label{sec:va-to-der_2D}
To make a GP prediction of a derivative value at $\bx_*$ using 2D volume average values as input, one extra step is required in the calculation of \cref{sec:va-to-pt_2D}. While we keep the $\mathcal{L}_1(K_{\text{SE}}(\bx,\bx'))$ kernel the same, we need to add the sought differential operation at $\bx_*$ to the $\mathcal{L}_2$ kernel operation in \cref{eq:va_2D_err} by taking respective derivatives of it. For example, to pursue $\partial_{xy}$ from the volume average input values, we define the kernel to be
%%%%%%%
\begin{align}
\mathcal{L}_2({K}_{\text{SE}}(\bx_*,\bx')) = 
&\prod_{d=x,y}
\frac{\partial}{\partial d}
\sqrt{\frac{\pi}{2}}\frac {\ell }{{\Delta d}} 
\left\{ 
\operatorname{erf} \Bigg( \frac{\big( \Delta_{d d'} + 1/2\big)}{\sqrt{2} \ell / \Delta d} \Bigg) - 
\operatorname{erf} \Bigg( \frac{\big( \Delta_{d d'} - 1/2\big)}{\sqrt{2} \ell / \Delta d} \Bigg) 
\right\} \Bigg|_{d=d_*}\nonumber
\\
=
&\prod_{d=x,y}
\frac {1 }{{\Delta d}} 
\left\{ 
\exp\Bigg(-\frac{(1+2\Delta_{d_*d'})^2}{8 (\ell /\Delta d)^2}\Bigg) - 
\exp\Bigg(-\frac{(1-2\Delta_{d_*d'})^2}{8 (\ell /\Delta d)^2}\Bigg)
\right\}.
\label{eq:va_to_der_2D}
\end{align}

In \cref{tab:Voleave_to_dxdy_2d}, we demonstrate the experimental convergence rates for the cross derivative approximations using the kernel in \cref{eq:va_to_der_2D}.
The anticipated convergence rate in this case is to be $(2r_{gp}-1)$ with the cross derivative as output. The results in \cref{tab:Voleave_to_dxdy_2d} show that the EOC rates are a little higher than  $(2r_{gp}-1)$ for each $r_{gp}=1,2,3$, which is also observed in the \textit{effective} order of convergence rates in \cref{tab:stencil_convergences} for the blocky diamond stencil.

%%%%%%%%%%%%%%%%%%%%%%%%%%%%%%%
%       Table 13
%%%%%%%%%%%%%%%%%%%%%%%%%%%%%%%
\begin{table}[ht!]
    %\footnotesize
    \scriptsize
        \centering
       \caption{Grid convergence results for the 2D volume-average-to-point conversion in \cref{sec:va-to-der_2D}. Input values are the integral average values $\frac{1}{\Delta x \Delta y}\int_{I_i} \int_{I_j}f(x,y) dxdy$ at cell centers $(x_i,y_j)$ and output values are the cross derivative $\partial_{xy} f(x_{i+1/2},y_{i+1/2})$  evaluated at cell corner locations for  $f(x,y) = e^{-2x}\sin(4 \pi y) + x e^{-y}\cos(2 \pi x) $ on $[0, 1]\times[0,1]$. GP utilizes the kernel in \cref{eq:va_to_der_2D} with $\ell =0.05$.}
        \begin{tabular}{@{}cccc|ccc|cccc@{}}
            \toprule
            \multirow{2}{*}{Grid Res.} & \multicolumn{3}{c}{$r_{gp} = 1$} & \multicolumn{3}{c}{$r_{gp} = 2$} & \multicolumn{3}{c}{$r_{gp} = 3$} \\ 
            \cmidrule(lr){2-4} \cmidrule(lr){5-7} \cmidrule(lr){8-10} 
            & $L_1$ error &  $L_2$ error & $L_\infty$ error & $L_1$ error &  $L_2$ error & $L_\infty$ error & $L_1$ error &  $L_2$ error & $L_\infty$ error \\
            \midrule
            \( 16 \)   & 8.7267e+00 & 1.0552e+01 & 2.6820e+01 & 5.8462e+00 & 6.9295e+00 & 1.6537e+01 & 1.5169e+00 & 2.0003e+00 & 6.4048e+00 \\
            \( 32 \)   & 1.5016e+00 & 2.0064e+00 & 7.9577e+00 & 5.1291e-01 & 6.3349e-01 & 2.1593e+00 & 1.3965e-01 & 1.8245e-01 & 6.2518e-01 \\
            \( 64 \)  & 1.4811e-01 & 2.6887e-01 & 1.2779e+00 & 1.6540e-02 & 3.0654e-02 & 1.6658e-01 & 8.8849e-04 & 1.6420e-03 & 8.6167e-03 \\
            \( 128 \)  & 5.0073e-02 & 6.9984e-02 & 3.5331e-01 & 7.6547e-04 & 1.1173e-03 & 6.7764e-03 & 1.7477e-05 & 2.5667e-05 & 1.4431e-04 \\
            \( 256 \)  & 1.6491e-02 & 2.1352e-02 & 7.7280e-02 & 7.1038e-05 & 9.2609e-05 & 3.6246e-04 & 3.9564e-07 & 5.1600e-07 & 1.9633e-06 \\

            \midrule
            & $L_1$ EOC &  $L_2$ EOC & $L_\infty$ EOC & $L_1$ EOC &  $L_2$ EOC & $L_\infty$ EOC & $L_1$ EOC &  $L_2$ EOC & $L_\infty$ EOC \\
            \midrule
            \( 16 \)   & -- & -- & -- & -- & -- & -- & -- & -- & -- \\
            \( 32 \)   & 2.54 & 2.39 & 1.75 & 3.51 & 3.45 & 2.94 & 3.44 & 3.45 & 3.36 \\
            \( 64 \)  & 3.34 & 2.90 & 2.64 & 4.95 & 4.37 & 3.70 & 7.30 & 6.80 & 6.18 \\
            \( 128 \)  & 1.56 & 1.94 & 1.85 & 4.43 & 4.78 & 4.62 & 5.67 & 6.00 & 5.90 \\
            \( 256 \)  & 1.60 & 1.71 & 2.19 & 3.43 & 3.59 & 4.22 & 5.47 & 5.64 & 6.20 \\
            \bottomrule
        \end{tabular} \label{tab:Voleave_to_dxdy_2d}
    \end{table}

%%%%%%%%%%%%%%%%%%%%%%%%%%%%%%%
%       Section 7
%%%%%%%%%%%%%%%%%%%%%%%%%%%%%%%
\section{DAS Kernel for Non-Oscillatory Approximation to Discontinuous Functions}\label{sec:gp_das}
Kernel selection is a crucial step in obtaining accurate and reliable outputs in GP models. As we have discussed so far for continuous function approximations, we can capture the underlying patterns and structures of target functions within the data by carefully choosing appropriate kernel functions. This, in turn, ensures that our model is capable of making accurate predictions and generalizing them well to unseen data points. The kernel in GP models not only serves as a mathematical function but also establishes a GP prior that expresses our assumptions about the underlying relationship in the data. By specifying a particular kernel, we are making a strong assumption about the smoothness, periodicity, and correlation structure of the function that is being modeled. Thus, the choice of a kernel directly impacts the resulting GP model's performance, and different types of data require different kernel choices.

In this section, we further extend our kernel investigation beyond smooth function approximations and explore appropriate kernel selection for discontinuous functions. Discrete approximation to discontinuous function data is one of the major research topics in numerical modeling of all times, where the main emphasis is dedicated to producing non-oscillatory solutions across sharp gradients. To demonstrate, let us consider a test problem of reconstructing pointwise data at a shock interface. We will analyze performances of three different kernels, including $K_{\text{SE}}$ in \cref{eq:se}, the so-called neural network kernel $K_{\text{NN}}$ \cite{rasmussen2005} given by
\begin{equation}
K_{\text{NN}}\left(\mathbf{x}, \mathbf{x}^{\prime}\right) =\frac{2}{\pi} \sin ^{-1}\left(\frac{2 \tilde{\mathbf{x}}^{\top} \Sigma \tilde{\mathbf{x}}^{\prime}}{\sqrt{\left(1+2 \tilde{\mathbf{x}}^{\top} \Sigma \tilde{\mathbf{x}}\right)\left(1+2 \tilde{\mathbf{x}}^{\prime \top} \Sigma \tilde{\mathbf{x}}^{\prime}\right)}}\right),
\label{eq:NN_kernel}
\end{equation}
and lastly our new ``Discontinous ArcSin kernel'' (or the DAS kernel for short) $K_{\text{DAS}}$ that takes of the form of
\begin{equation}
K_{\text{DAS}}\left(\mathbf{x}, \mathbf{x}^{\prime}\right) =\frac{2}{\pi} \sin ^{-1}\left(\frac{ \exp \left( - \sqrt{(\mathbf{x} - \mathbf{x^\prime})^\top (\mathbf{x} - \mathbf{x^\prime})} \right) }{\sqrt{  \prod_{i=1}^d\left(1 + 2  ( 1  + \  {(x_i - x_i^\prime )^2}\right) )}}\right).
\label{eq:DAS_kernel}
\end{equation}
In \cref{eq:NN_kernel}, $\tilde{\mathbf{x}}=(1,\bx)^T = \left(1, x_1, \ldots, x_d\right)^{\top}$ is an augmented vector of $\bx$ with $d$ being the number of spatial dimensions, which must also be the size of $\bx$ and $\bx'$. We also set $\Sigma={\text{diag}}(\sigma_0^2,\sigma^2)$ following \cite{rasmussen2005}. The augmentation with unity prevents the kernel from failing to be symmetric semi-positive definite--a requirement for GP kernel functions--when $\bx=\bx'=\mathbf{0}$ especially in the diagonal entries, keeping its eigenvalues to be non-negative. The functions sampled from this NN kernel can be viewed as superpositions of error functions, $\text{erf}(u_0+ux)$, where $\sigma_0^2$ controls the variance of $u_0$ while $\sigma^2$ controls $u$, in which case they respectively influence how much these sample functions are offset from the origin and their scaling on the $x$-axis \cite{rasmussen2005}.

Below, we discuss how these three kernels are compared on data predictions across sharp discontinuities, particularly emphasizing the improved non-oscillatory shock-capturing abilities of the new DAS kernel. 
First, we remark that the SE and DAS kernels are a \textit{stationary} kernel, which means it only depends on the `distance' between observed input data points at $\bx$ and $\bx'$ rather than the actual values of them. That is, SE and DAS are invariant to translations in the input space. This is computationally efficient because $K^{-1}$ only needs to be computed once, saved, and reused over computational steps as long as the computational grid configuration remains static. This is not the case for the NN kernel. It is \textit{non-stationary}; hence its inverse matrix must to be computed at each input location. The non-stationary property of $K_{\text{NN}}$ makes it computationally challenging even for small problems, not to mention big problems with increasing grid resolutions. 

%%%%%%%%%%%%%%%%%%%%%%%%%%%%%%%
%       Fig. 4
%%%%%%%%%%%%%%%%%%%%%%%%%%%%%%%
\begin{figure}[th!]
  \centering
  \includegraphics[width=0.8\textwidth]{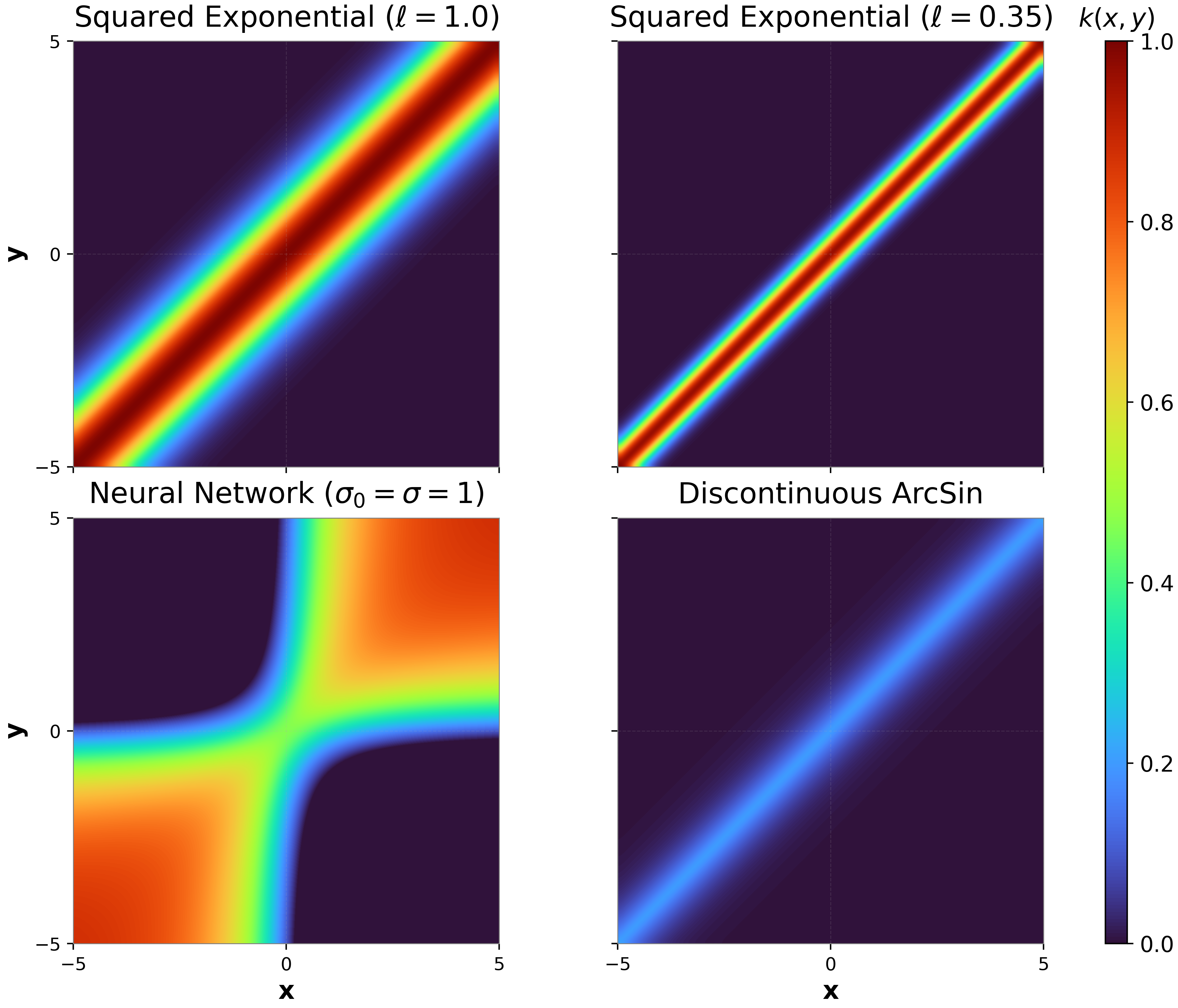} 
    \caption{Heat map comparison of kernels. \textbf{(Top left)} SE kernel with $\ell = 1.0$. \textbf{(Top right)} SE kernel with $\ell = 0.35$. \textbf{(Bottom left)} NN kernel with $\sigma = \sigma_0 = 1$. \textbf{(Bottom right)} DAS kernel.}
  \label{fig:heat_map_kernels}
\end{figure}

The above observations motivate us to design $K_{\text{DAS}}$ to be a stationary kernel. In \cref{fig:heat_map_kernels}, we show the kernel heat maps. The way to read this map is by considering each point $\bx=(x,y)$ as an input to the kernel. The color scale represents the value of the kernel: the higher the value, the larger the weight or importance the GP will place on that point. Noticeably, we see that $K_{\text{DAS}}$ takes the maximum value along a very thin diagonal band. The consequence is $K_{\text{DAS}}$'s strong prediction between the data points separated by, namely, the bandwidth, $\delta_b$, where it is seen that $\delta_b \sim \mathcal{O}(\Delta m)$ with $\Delta m = \max_{d=x,y}\{\Delta d\}$. On the other hand, if any data points are separated by a distance larger than $\delta_b$, i.e., 10 or more cells away from each other, the correlation of them rapidly drops with the increasing inter-distance and $K_{\text{DAS}}$'s predictions become insignificant. By design, henceforth, $K_{\text{DAS}}$ is apt to work very well for narrow fluid structures such as shocks or discontinuities. Also importantly, for the same data values at different grid locations, $K_{\text{DAS}}$'s stationary property returns the same kernel value as long as the distance between the data is the same. This is an important consistency concern for numerical methods.

Contrary to the well-confined thin-band structure in the stationary $K_{\text{DAS}}$ kernel, the non-stationary effects in the NN kernel are evident in the heat map with an increased area of high kernel values. With $K_{\text{NN}}$, GP correlations depend on the actual values of $\tilde{\mathbf{x}}$ and $\tilde{\mathbf{x}}'$, thereby the heat map appears to be widely spread out across the large area of the entire domain. It is true that the NN kernel can be tuned to produce a more confined, narrow band-like structure by increasing $\sigma$ from the current value of 1 in \cref{fig:heat_map_kernels}. Nevertheless, one still faces the non-stationary effects, including the need for inverting the kernel at each input location of $\bx$ and $\bx'$. We present $K_{\text{NN}}$'s heat map using $\sigma_0=\sigma=1$ to visualize the non-stationary effects in a clear way.

%%%%%%%%%%%%%%%%%%%%%%%%%%%%%%%
%       Table 14
%%%%%%%%%%%%%%%%%%%%%%%%%%%%%%%
\begin{table}[ht!]
    %\footnotesize
    \scriptsize
    \centering
    \caption{Grid convergence results for the 1D baseline point-to-point interpolation using the NN kernel in \cref{eq:NN_kernel}. The NN prediction takes cell center function values $f(x_i)$ as input for $f(x) = e^{-x}\sin(4\pi x) \cos(2\pi x)$ on $[0, 1]$ and computes $f(x_{i+1/2})$ using $\sigma = \sigma_0 = 1$.}
    \begin{tabular}{@{}cccc|ccc|cccc@{}}
        \toprule
        \multirow{2}{*}{Grid Res.} & \multicolumn{3}{c}{$r_{gp} = 1$} & \multicolumn{3}{c}{$r_{gp} = 2$} & \multicolumn{3}{c}{$r_{gp} = 3$} \\ 
        \cmidrule(lr){2-4} \cmidrule(lr){5-7} \cmidrule(lr){8-10} 
        & $L_1$ error &  $L_2$ error & $L_\infty$ error & $L_1$ error &  $L_2$ error & $L_\infty$ error & $L_1$ error &  $L_2$ error & $L_\infty$ error \\
        \midrule
        \( 16 \)   & 2.0485e-02 & 2.4082e-02 & 4.5616e-02 & 5.0616e-03 & 5.8995e-03 & 1.1189e-02 & 1.1641e-03 & 1.3252e-03 & 2.3945e-03 \\
        \( 32 \)   & 2.6048e-03 & 2.9700e-03 & 5.1987e-03 & 1.5872e-04 & 1.8007e-04 & 3.2454e-04 & 8.8349e-06 & 9.7185e-06 & 1.6531e-05 \\
        \( 64 \)  & 3.2722e-04 & 3.7604e-04 & 7.4039e-04 & 4.8296e-06 & 5.5502e-06 & 1.0466e-05 & 6.3433e-08 & 7.1688e-08 & 1.2562e-07 \\
        \( 128 \)  & 4.0433e-05 & 4.6538e-05 & 9.8315e-05 & 1.4945e-07 & 1.7071e-07 & 3.3212e-07 & 4.7695e-10 & 5.3743e-10 & 9.4100e-10 \\
        \( 256 \)  & 5.0100e-06 & 5.7736e-06 & 1.2506e-05 & 4.5900e-09 & 5.2572e-09 & 1.0579e-08 & 4.3487e-12 & 5.2229e-12 & 1.5137e-11 \\
        \midrule
        & $L_1$ EOC &  $L_2$ EOC & $L_\infty$ EOC & $L_1$ EOC &  $L_2$ EOC & $L_\infty$ EOC & $L_1$ EOC &  $L_2$ EOC & $L_\infty$ EOC \\
        \midrule
        \( 16 \)   & -- & -- & -- & -- & -- & -- & -- & -- & -- \\
        \( 32 \)   & 2.975 & 3.019 & 3.133 & 4.995 & 5.034 & 5.108 & 7.042 & 7.091 & 7.178 \\
        \( 64 \)  & 2.993 & 2.981 & 2.812 & 5.038 & 5.020 & 4.955 & 7.122 & 7.083 & 7.040 \\
        \( 128 \)  & 3.017 & 3.014 & 2.913 & 5.014 & 5.023 & 4.978 & 7.055 & 7.059 & 7.061 \\
        \( 256 \)  & 3.013 & 3.011 & 2.975 & 5.025 & 5.021 & 4.972 & 6.777 & 6.685 & 5.958 \\

        \bottomrule
    \end{tabular}    \label{tab:Pt_to_Pt_1D_NN}
\end{table}

%%%%%%%%%%%%%%%%%%%%%%%%%%%%%%%
%       Table 15
%%%%%%%%%%%%%%%%%%%%%%%%%%%%%%%
\begin{table}[ht!]
    %\footnotesize
    \scriptsize
    \centering
    \caption{Grid convergence results for the 1D baseline point-to-point interpolation using the DAS kernel in \cref{eq:DAS_kernel}. The DAS prediction takes cell center function values $f(x_i)$ as input for $f(x) = e^{-x}\sin(4\pi x) \cos(2\pi x)$ on $[0, 1]$ and computes $f(x_{i+1/2})$.}
    \begin{tabular}{@{}cccc|ccc|cccc@{}}
        \toprule
        \multirow{2}{*}{Grid Res.} & \multicolumn{3}{c}{$r_{gp} = 1$} & \multicolumn{3}{c}{$r_{gp} = 2$} & \multicolumn{3}{c}{$r_{gp} = 3$} \\ 
        \cmidrule(lr){2-4} \cmidrule(lr){5-7} \cmidrule(lr){8-10} 
        & $L_1$ error &  $L_2$ error & $L_\infty$ error & $L_1$ error &  $L_2$ error & $L_\infty$ error & $L_1$ error &  $L_2$ error & $L_\infty$ error \\
        \midrule
        \( 16 \)   & 3.5577e-02 & 4.2384e-02 & 8.6919e-02 & 3.5273e-02 & 4.2029e-02 & 8.5971e-02 & 3.5243e-02 & 4.2033e-02 & 8.5812e-02 \\
        \( 32 \)   & 9.1863e-03 & 1.0765e-02 & 2.2903e-02 & 9.1511e-03 & 1.0725e-02 & 2.2820e-02 & 9.1060e-03 & 1.0675e-02 & 2.2742e-02 \\
        \( 64 \)  & 2.2622e-03 & 2.6473e-03 & 5.5801e-03 & 2.2603e-03 & 2.6453e-03 & 5.5753e-03 & 2.2566e-03 & 2.6409e-03 & 5.5670e-03 \\
        \( 128 \)  & 5.6063e-04 & 6.5374e-04 & 1.3808e-03 & 5.6053e-04 & 6.5363e-04 & 1.3805e-03 & 5.6032e-04 & 6.5338e-04 & 1.3801e-03 \\
        \( 256 \)  & 1.3931e-04 & 1.6230e-04 & 3.4265e-04 & 1.3930e-04 & 1.6229e-04 & 3.4264e-04 & 1.3929e-04 & 1.6228e-04 & 3.4262e-04 \\
        \midrule
        & $L_1$ EOC &  $L_2$ EOC & $L_\infty$ EOC & $L_1$ EOC &  $L_2$ EOC & $L_\infty$ EOC & $L_1$ EOC &  $L_2$ EOC & $L_\infty$ EOC \\
        \midrule
        \( 16 \)   & -- & -- & -- & -- & -- & -- & -- & -- & -- \\
        \( 32 \)   & 1.953 & 1.977 & 1.924 & 1.947 & 1.970 & 1.914 & 1.952 & 1.977 & 1.916 \\
        \( 64 \)  & 2.022 & 2.024 & 2.037 & 2.017 & 2.019 & 2.033 & 2.013 & 2.015 & 2.030 \\
        \( 128 \)  & 2.013 & 2.018 & 2.015 & 2.012 & 2.017 & 2.014 & 2.010 & 2.015 & 2.012 \\
        \( 256 \)  & 2.009 & 2.010 & 2.011 & 2.009 & 2.010 & 2.010 & 2.008 & 2.009 & 2.010 \\

        \bottomrule
    \end{tabular}\label{tab:Pt_to_Pt_1D_DAS}
\end{table}

\begin{figure}[h!]
  \centering
  \includegraphics[width=0.8\textwidth]{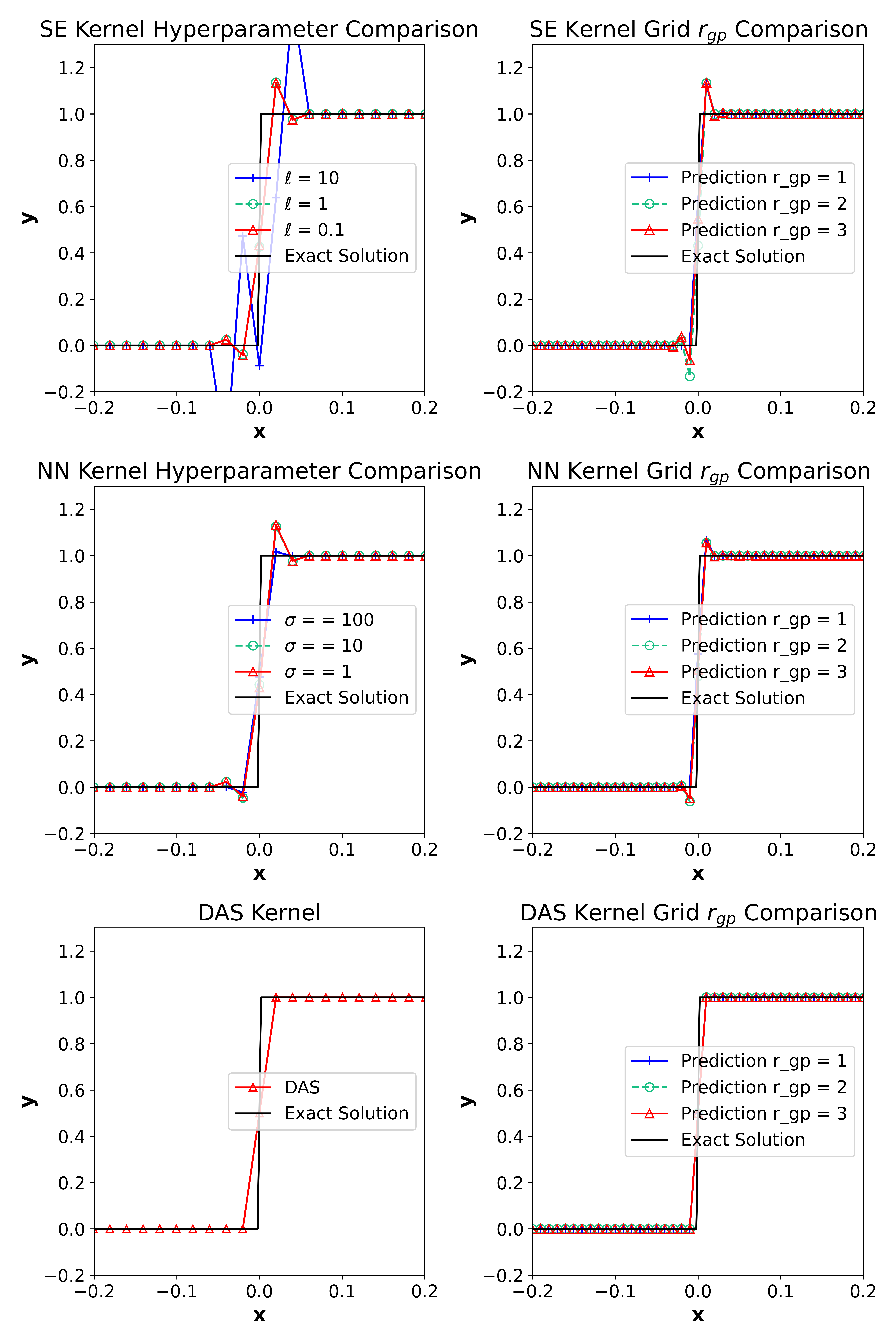}  
  \caption{We test the problem of interpolating across a single discontinuity using our GP point-to-point interpolation and compare different kernels. The solution is computed over $[-0.2, 0.2]$ with $N_x = 20$ for the left panel and $N_x = 40$ for the right. \textbf{(Left)} The hyperparameters of the kernels are varied on each kernel holding $r_{gp} = 2$. The SE kernel is varied over $\ell$ values of 0.1, 1.0, and 10. The NN kernel is varied over $\sigma$ values of 1, 10, and 100. We keep $\sigma_0 = 1$ as it has little effect on the solution. DAS does not have any hyperparameters and is held at constant. \textbf{(Right)} The stencil order is varied from $r_{gp} = 1,2,3$ for each kernel choice. The SE kernel utilizes $\ell = 0.1$ and the NN kernel uses $\sigma = 100$ and $\sigma_0 = 1$.
  }
  \label{fig:kernel_shock}
\end{figure}

\begin{figure}[h!]
  \centering
  \includegraphics[width=0.95\textwidth]{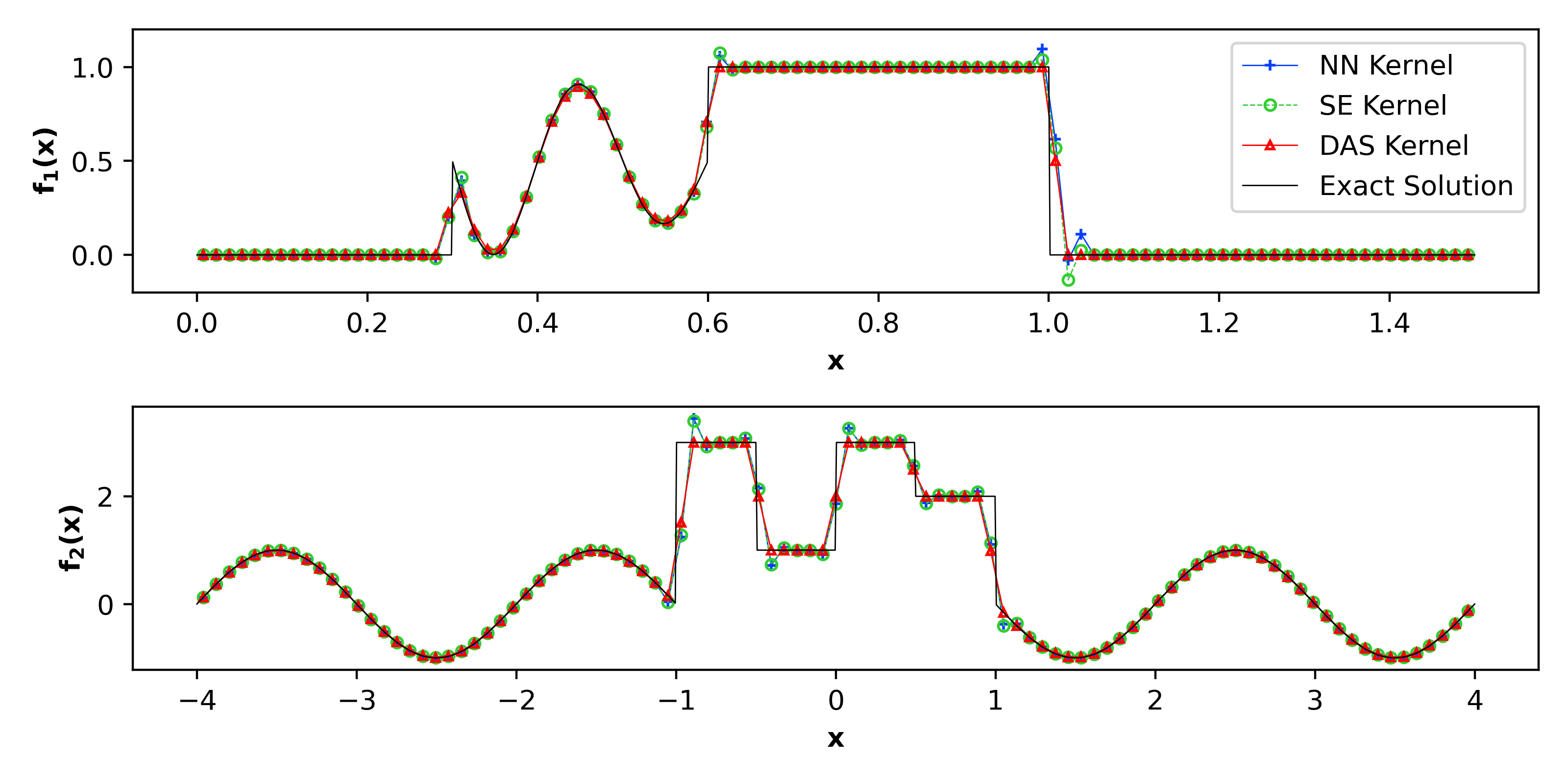}  
\caption{Comparison of point-to-point interpolation on three different profiles using the SE, NN, and DAS kernels. The NN kernel uses $\sigma=\sigma_0=1$, and the SE kernel uses $\ell = 12 \Delta x$.
\textbf{(Top)} The function, $f_1$, is a piecewise function defined on $[0, 1.5]$ by $f_1(x)=0$ for $0 \leq x < 0.3$ and $1.0 \leq x \leq 1.5$; $f_1(x) = e^{-2x} \sin(10\pi x) + 1/2$ for $0.3 \leq x < 0.6$; $f_1(x)=1$ for $0.6 \leq x < 1.0$. \textbf{(Bottom)} The function, $f_2$, is a piecewise function defined on $[-4,4]$ by $f_2(x) = \sin(\pi x)$ for $|x| \geq 1.0$; $f_2(x)=3$ for $-1.0 < x \leq -0.5$ and $0 < x \leq 0.5$; $f_2(x)=1$ for $-0.5 < x \leq 0.0$; $f_2(x)=2$ for $0.5 < x \leq 1.0$. All GP calculations were performed using $N_x = 100$ grid resolutions on each domain with $r_{gp} = 2$. The reference exact solution is shown using $N_x=1000$.}
  \label{fig:compound_waves}
\end{figure}

First, we examine the convergence of the three kernels on smooth data. For the baseline pointwise interpolation, both $K_{\text{SE}}$ and $K_{\text{NN}}$ feature a rate of convergence order $(2r_{gp} + 1)$ on smooth data as shown in \cref{tab:pt-to-pt_1d,tab:pt-to-pt_2d} and \cref{tab:Pt_to_Pt_1D_NN}, respectively. The $K_{\text{DAS}}$ lags behind significantly in this regard and demonstrates a fixed second order convergence rate regardless of stencil sizes as shown in \cref{tab:Pt_to_Pt_1D_DAS}. Despite its inferior convergence rate on smooth flows, the true value of the DAS kernel is to be emphasized and highlighted in relation to discontinuous flow problems. In \cref{fig:kernel_shock,fig:compound_waves}, we present how these three kernels perform in approximating data predictions across shocks and discontinuities for interpolation.

%%%%%%%%%%%%%%%%%%%%%%%%%%%%%%%
%       Fig. 5
%%%%%%%%%%%%%%%%%%%%%%%%%%%%%%%

%%%%%%%%%%%%%%%%%%%%%%%%%%%%%%%
%       Fig. 6
%%%%%%%%%%%%%%%%%%%%%%%%%%%%%%%

In the first four panels of \cref{fig:kernel_shock} (from left to right and top to bottom), we see that $K_{\text{NN}}$ performs better than $K_{\text{SE}}$ showing a decreased amount of oscillations at the discontinuity. Nonetheless, the shock oscillations persist in both kernels for all three choices of GP radius, $r_{gp} = 1, 2, 3$, and the level of oscillations is not seen to be controlled by varying the size of the GP stencil, i.e., irrelevant to modeling accuracy of the two kernels. As discussed, $K_{\text{NN}}$ has two tunable parameters, $\sigma$ and $\sigma_0$. We find that increasing $\sigma$ helps improve its performance for shock capturing but does not completely remove shock oscillations.

In contrast, our new $K_{\text{DAS}}$ kernel in the bottom two panels shows very promising results with no grid-scale oscillations at the jump interface. The DAS kernel's second-order insensitivity to $r_{gp}$ is also observed on the bottom right panel, where all three calculations with different radii behave nearly the same way. 

In \cref{fig:compound_waves}, the shock-capturing performance of the three kernels is further compared in two additional compound wave configurations. Again, the DAS kernel outperforms the NN and SE kernels at the discontinuities, displaying no observable oscillations, while all three kernels behave equally well on the smooth parts. The results in \cref{fig:kernel_shock,fig:compound_waves} suggest that $K_{\text{DAS}}$ can be used in a shock-capturing code selectively at shocks and discontinuities within a similar framework of GP-MOOD \cite{bourgeois2022gp} as part of hybridizing non-oscillatory stable second-order GP solutions with DAS near shocks/discontinuities while high-order GP solutions with SE elsewhere. This topic will be investigated in our future work.

\begin{figure}[h!]
  \centering
  \includegraphics[width=1\textwidth]{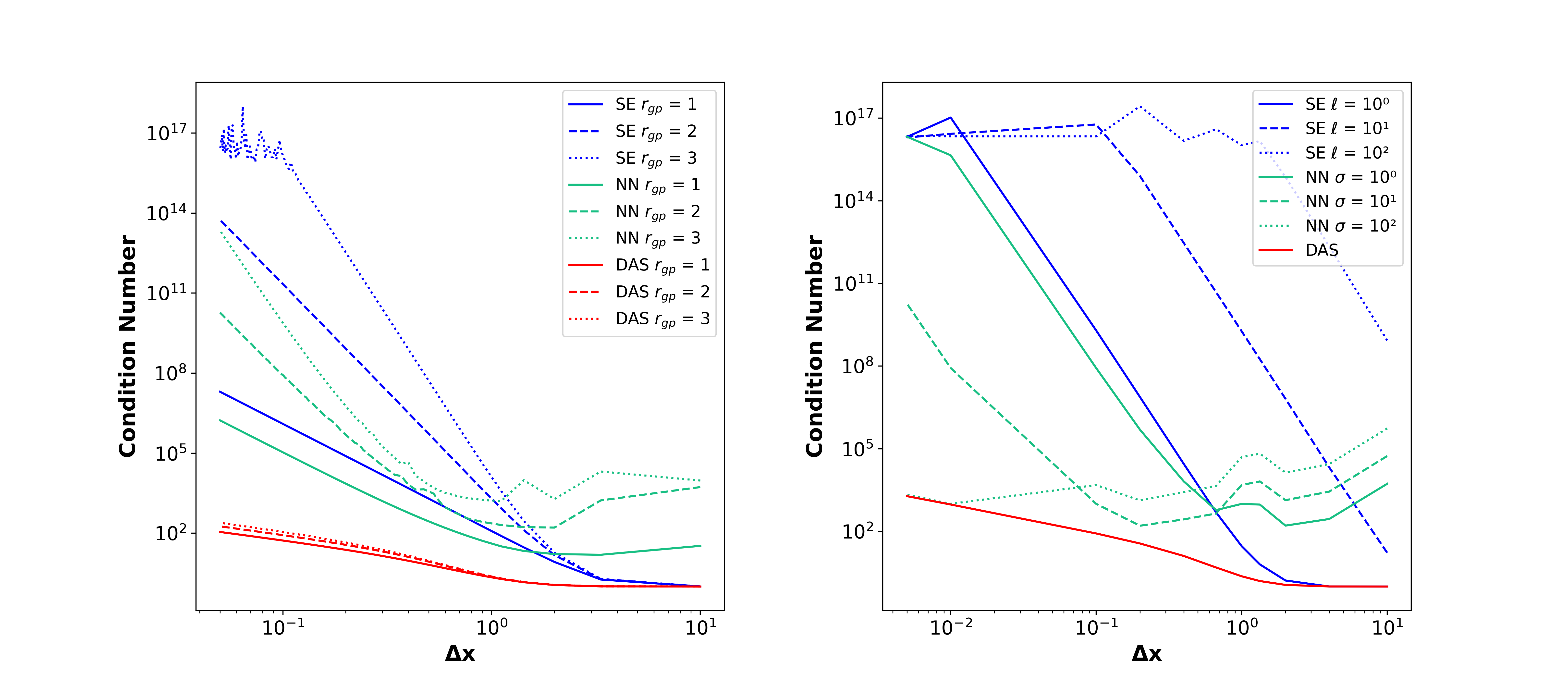}  
  \caption{Condition number of $\mathbf{K}$ plotted as a function of grid spacing $\Delta x$.
  \textbf{(Left)} Comparison of the three kernels for various stencil sizes. The hyperparameters are held at $\ell = 1.8$ for SE and $\sigma = \sigma_0  =1$ for NN, while $r_{gp}$ is varied for all cases. 
  \textbf{(Right)} Comparison of the three kernels for hyperparameter values. We set $r_{gp} = 2$ for all kernels and hyperparameters are varied according to the legend.}
  \label{fig:condition_number}
\end{figure}

In \cref{fig:condition_number}, we compare the condition number of the covariance kernel matrices using the three kernel functions, $K_\text{SE}$, $K_\text{NN}$, and $K_\text{DAS}$. On the left, we examine the condition number of each kernel matrix as a function of grid spacing $\Delta x$ across different stencil sizes. On the right, we compare how the condition number varies with $\Delta x$ for different values of the hyperparameters, e.g., $\ell$ for the SE kernel and $\sigma$ for the NN kernel. 
Our analysis reveals that the choices of $\ell$ for $K_\text{SE}$ and $\sigma$ for $K_\text{NN}$ dramatically impact the condition number, with the potential for extremely high values. It is always desirable to keep condition numbers as low as possible in numerical calculations to prevent the corresponding kernel matrix from being singular. Compared to such sensitive behaviors in $K_\text{SE}$ and $K_\text{NN}$, the $K_\text{DAS}$ kernel demonstrates consistently low condition numbers less than $10^5$ over the wide range of grid resolutions up to $10^4$. The low condition number with $K_\text{DAS}$ assures that its relevant calculations are highly reliable on a wide spectrum of discrete grid scales without experiencing numerical issues due to singularity.

It is evident in \cref{fig:kernel_shock,fig:compound_waves} that the discontinuous data-fitting with DAS is oscillation-free, tremendously outperforming the SE and NN kernels. There are two main reasons for the DAS kernel's excellent performance on interpolation of discontinuous data. The first is the small region of dependence seen in \cref{fig:heat_map_kernels}. Regardless of $r_{gp}$, the DAS kernel maintains this narrow band, which leads to a lower order solution, in this case, second order. The second reason is the low condition number property. We can explain this by noting that the discontinuous data-fitting problem is a dot product operation of the two vectors $\bz_*^T$ and $\bff$ in \cref{eq:PostMean} from the mathematical perspective. %Here, we safely assume that the data vector $\bff$ contains non-oscillatory exact function values $f(x_i) = f_i$ sampled from a GP stencil. As such, t
The higher the condition number of $\bK$, the more the $\bz_*^T$ becomes ill-conditioned, which negatively impacts the accuracy and stability of the data-fitting problem. Equivalently, $\bz_*^T$ becomes well-conditioned for the kernel matrices with low condition numbers. This relationship between the condition number and the discontinuous data-fitting performance is observed in the results of \cref{fig:kernel_shock,fig:condition_number} too. In the right panel of \cref{fig:condition_number}, we see that, among the various kernels under comparison, the condition numbers of the NN kernel with $\sigma=100$ and the DAS kernel are the closest to each other over a wide range of grid resolutions. Therefore, we can expect that the data-fitting performance of NN would be similar to that of DAS. Indeed, in the middle left panel in \cref{fig:kernel_shock}, the NN kernel with $\sigma=100$ produces the least oscillatory interpolation across the discontinuity, demonstrating that its solution is qualitatively comparable to the DAS solution in the bottom left panel.

%%%%%%%%%%%%%%%%%%%%%%%%%%%%%%%
%       Fig. 7
%%%%%%%%%%%%%%%%%%%%%%%%%%%%%%%

%%%%%%%%%%%%%%%%%%%%%%%%%%%%%%%
%       Conclusion
%%%%%%%%%%%%%%%%%%%%%%%%%%%%%%%
\section{Conclusion}
We have presented a kernel-based GP approach to approximating function values under various linear operations. These operations involve function interpolations, numerical derivatives and integrations, and input-output data type conversions, all appearing very commonly in a broad spectrum of science and engineering applications. In particular, these operations are crucial building blocks in numerical modeling across popular discrete fluid dynamics methods, including finite difference, finite volume, finite element, Galerkin, and many others. This study showed that the GP approach can deliver a variable order of convergence depending on the GP stencil size determined by the GP radius, $r_{gp}$. Instead of conventional polynomial-based methods, these GP methods can be readily adopted in the existing differential, integral, interpolation, and reconstruction data fitting problems.

An approach to integrate the Taylor series analysis into the GP modeling was newly introduced to measure the effective order of convergence in multi-dimensional GP stencils. This tool is a convenient predictive metric for selecting the optimal GP stencil among various configurations, analogous to the leading error analysis in the conventional polynomial-based modeling paradigm. In addition, the Taylor series-based GP analysis provides GP modeling with a more mathematically rigorous and systematic toolset for quantifying discrete errors in numerical approximations.

We also introduced a new stationary DAS kernel for discontinuous data fitting. Compared to the SE and NN kernels, the DAS matrix features relatively low condition numbers in a wide range of length scales with a narrow region of dependence, directly influencing its non-oscillatory performance in discontinuous data-fitting problems. This novel feature in DAS can provide another avenue for designing shock-capturing numerical methods for compressible flows, which we will investigate in the future.

%%%%%%%%%%%%%%%%%%%%%%%%%%%%%%%
%       Acknowledgements
%%%%%%%%%%%%%%%%%%%%%%%%%%%%%%%
\section*{Acknowledgements}\label{sec:ack}
This work was supported in part by the National Science Foundation under grants AST-1908834 and AST-2307684. 
Both authors would like to thank our friends and colleagues, Drs. Ian May, Youngjun Lee, and Carlo Graziani, for their valuable contributions and consultations in the early phases of this work.

%
%%%%%%%%%%%%%%%%%%%%%%%%%%%%%%%
%       Section Appendix
%%%%%%%%%%%%%%%%%%%%%%%%%%%%%%%
\appendix

%%%%%%%%%%%%%%%%%%%%%%%%
%       Subsection - Variance of the Posterior
%%%%%%%%%%%%%%%%%%%%%%%%
\section{Variance of the Posterior}
\label{apdx:variance}

\begin{figure}[h!]
  \centering
  \includegraphics[width=0.8\textwidth]{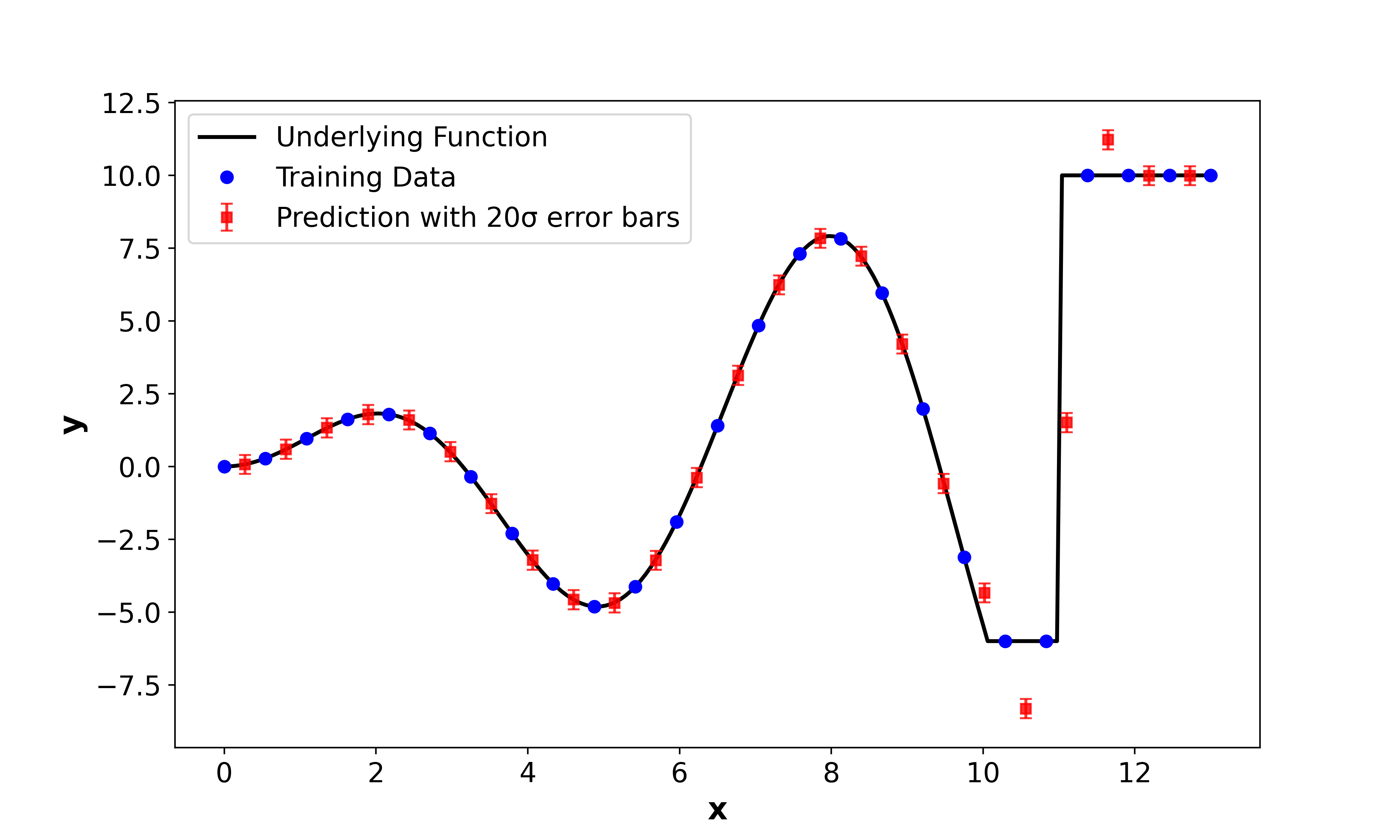}  
  \caption{Gaussian Process interpolation is applied to a piecewise function. The function is defined as $f(x) = x\sin(x)$ for all $x<10$, $f(x)=-6$ for all $10 \leq x < 11.5$ and $f(x) = 10$ for all $x \geq 11.5$ on a domain $[0,13]$ with 25 cells. For the GP calculation, we use the SE kernel with $r_{gp} = 1$ and $\ell = 0.8$. Since the variance of the posterior calculation only depends on the distance between two points with the SE kernel, this does not give any useful information that can be used to inform sharp-gradient-aware reconstruction/interpolation of the solution near shocks or discontinuities.}
  \label{fig:ci_plot}
\end{figure}

As previously described, the variance of the posterior provides a measure of the confidence in predictions made by our Gaussian Process (GP) model. In \cref{fig:ci_plot}, we apply the model to interpolate a piecewise function using a squared exponential (SE) kernel, which explicitly depends on the Euclidean distance between prediction points ($x_*$) and training data locations ($x$). In our analysis, the posterior variance is thus determined exclusively by grid spacing and the chosen kernel parameters, independent of the actual input data values. While this distance-based measure is straightforward, it may not justify the computational effort required, particularly since simpler distance metrics could provide similar insights in CFD applications. Other studies \cite{10.1145/1102351.1102413,10.1145/1273496.1273546} have shown that scaling the uncertainty measure by local derivatives or gradients of the data significantly improves its informativeness, particularly when capturing rapidly varying features. This might be useful in a region where the GP is used on CFD data with shocks where a high gradient in the interpolation would give a larger confidence interval. Although we explored alternative kernels in this work, these additional complexities provided limited practical value for our current low-dimensional context. Such kernels are advantageous in high-dimensional problems where simple distance metrics become insufficient. Future work will focus on incorporating data-driven derivative scaling into uncertainty estimation.

%%%%%%%%%%%%%%%%%%%%%%%%
%       Subsection - GP is closed under Linear 
%%%%%%%%%%%%%%%%%%%%%%%%
\section{GP is closed under linear operators}
\label{apdx:GP_closedLinear}

To prove \cref{eq:GP_under_linear}, we firstly show that
\beq
\mathbb{E} \left[ \mathcal{L}\Big(f(\mathbf{x})\big) \right] = 
\mathcal{L}\big( \mathbb{E} \left[f(\mathbf{x})\right]\big) =  
\mathcal{L}\big(m(\bx)\Big), 
\eeq
which proves that the mean of a linearly transformed function, $\mathcal{L}(f(\bx))$,
is the mean function of $f$ transformed by $\mathcal{L}$.
Secondly, for the covariance function, we show
\begin{align}
&\mathbb{E} 
\left[ 
\Big( 
\mathcal{L}_{\bx} \big( f(\bx) \big) - 
\mathbb{E} \left[ \mathcal{L}_{\bx}  \big(f(\bx) \big) \right]
\Big) 
\Big( 
\mathcal{L}_{\bx'} \big( f(\bx') \big) - 
\mathbb{E} \left[ \mathcal{L}_{\bx'}  \big(f(\bx') \big) \right]
\Big) 
\right] \\
&=
\mathbb{E} 
\left[ 
\Big( 
\mathcal{L}_\bx \big( f(\bx) - 
\mathbb{E} \left[  f(\bx) \right]\big)
\Big) 
\Big( 
\mathcal{L}_{\bx'} \big(f(\bx')  - 
\mathbb{E} \left[  f(\bx')  \right]\big)
\Big) 
\right] \\
&=
\mathcal{L}_\bx
\Bigg(
\mathcal{L}_{\bx'}
\Big(
\mathbb{E} 
\left[ 
\big( f(\bx) - m(\bx)\big) 
\big( f(\bx')  - m(\bx')\big) 
\right] 
\Big)
\Bigg) \\
&=
\mathcal{L}
\Big(
K(\bx,\bx')
\Big),
\end{align}
which proves that the covariance function for a linearly transformed function, $\mathcal{L}(f(\bx))$, is the
covariance function of $f$ transformed by $\mathcal{L}$.

%%%%%%%%%%%%%%%%%%%%%%%%
%       References
%%%%%%%%%%%%%%%%%%%%%%%%
\bibliography{refs_merged_new}
\end{document}